\documentclass[twocolumn]{aastex631}

\usepackage{newtxtext,newtxmath}
\usepackage{amsmath}
\usepackage{savesym}
\usepackage[T1]{fontenc}
\usepackage{xcolor}
\usepackage{graphics, graphicx}	
\usepackage{amsmath}	
\newcommand{\mr}{\mathrm}

\newcommand{\be}{\[\begin{aligned}}
\newcommand{\ee}{\end{aligned}\]}
\newcommand{\beq}{\begin{equation}}
\newcommand{\eeq}{\end{equation}}
\newcommand{\sr}{\mr{src}}
\newcommand{\ls}{\mr{lens}}

\begin{document}

\title{A Unified Model of Cosmic Ray Propagation and Radio Extreme Scattering Events from Intermittent Interstellar Structures}
\shorttitle{CR scattering and ESEs}
\shortauthors{Kempski, Li et al.}

\author{Philipp Kempski}
\altaffiliation{Philipp Kempski and Dongzi Li contributed equally and are co-lead authors.}
\affiliation{Department of Astrophysical Sciences, Princeton University, Princeton, NJ 08544, USA}
\author[0000-0001-7931-0607]{Dongzi Li}
\altaffiliation{Philipp Kempski and Dongzi Li contributed equally and are co-lead authors.}
\affiliation{Department of Astrophysical Sciences, Princeton University, Princeton, NJ 08544, USA}

\author[0000-0003-3806-8548]{Drummond B. Fielding}
\affiliation{Department of Astronomy, Cornell University, Ithaca, NY 14853, USA}

\author{Eliot Quataert}
\affiliation{Department of Astrophysical Sciences, Princeton University, Princeton, NJ 08544, USA}

\author[0000-0002-9656-4032]{E. Sterl Phinney}
\affiliation{Theoretical Astrophysics,
MC 350-17, California Institute of Technology,
Pasadena, CA 91125,  USA}

\author[0000-0003-1676-6126]{Matthew W. Kunz}
\affiliation{Department of Astrophysical Sciences, Princeton University, Princeton, NJ 08544, USA}
\affiliation{Princeton Plasma Physics Laboratory, PO Box 451, Princeton, NJ 08543, USA}

\author[0000-0003-3236-8769]{Dylan L. Jow}
\affiliation{Kavli Institute for Particle Astrophysics \& Cosmology, P.O. Box 2450, Stanford University, Stanford, CA 94305, USA}

\author[0000-0001-7801-0362]{Alexander A. Philippov}
\affiliation{Department of Physics, University of Maryland, College Park, MD 20742, USA}

\correspondingauthor{Philipp Kempski}
\email{pkempski@princeton.edu}
\correspondingauthor{Dongzi Li}
\email{dongzili@princeton.edu}



\begin{abstract}

Intermittent magnetic structures are a plausible candidate for explaining cosmic-ray (CR) diffusion rates derived from observed CR energy spectra. Independently, studies of extreme scattering events (ESEs) of radio quasars and pulsar scintillation have hinted that very straight, large-aspect-ratio, magnetic current sheets may be responsible for the localized large scattering of radio waves. The required shortest axis of the typical structures producing ESEs is of the same scale ($\sim$AU) as the gyroradii of $\sim$GeV CRs. In this paper, we propose that the same magnetic/density sheets can produce large scattering of both CRs and radio waves. We demonstrate that the geometry and volume filling factor of the sheets derived from quasar ESEs can explain the observed mean free path of GeV CRs without introducing free parameters. The model places constraints on the sheet geometry, such as straightness and large aspect ratio, and assumes the statistics of the sheets are similar throughout the Galactic volume. We, therefore, discuss observational tests of the sheet model, which includes observations of echoes in pulsars and fast radio bursts, gravitationally lensed quasars, the distribution of ESE durations, and spatial correlations between ESE events and rotation-measure fluctuations. Such tests will be enabled by upcoming wide-field radio instruments, including Canadian Hydrogen Observatory and Radio-transient Detector (CHORD) and Deep Synoptic Array 2000 Antennas (DSA-2000).
\end{abstract}

\keywords{Cosmic rays  --- Interstellar magnetic fields --- Interstellar medium --- Interstellar scattering --- Radio continuum emission}

\section{Introduction}
Cosmic rays (CRs) injected into the interstellar medium (ISM) of galaxies by supernovae and active galactic nuclei play an important role in galaxy formation, from the ionization of molecular clouds (\citealt{Dalgarno:2006}), to the heating of galactic halos (\citealt{Guo:2008}; \citealt{jp_1}; \citealt{jp_2}), to the launching of galactic outflows (\citealt{Ipavich:1975}; \citealt{Ruszkowski:2017}; \citealt{Hopkins:2021CRWIND}; \citealt{Thomas:2022}). However, the impact that cosmic rays have on their host systems in theoretical and numerical models depends sensitively on the assumed transport model (\citealt{Wiener:2017}; \citealt{Butsky:2018}; \citealt{Quataert:2022b}; \citealt{Quataert:2022}). Over the last several decades, the standard paradigm of CR transport has been that these high-energy particles are scattered by volume-filling small-amplitude magnetic fluctuations that permeate the ISM and halos of galaxies, either excited by the CR streaming instability (\citealt{KulsrudPearce:1969}; \citealt{Skilling:1971}; \citealt{Zweibel:2013}; \citealt{Zweibel:2017}) or due to extrinsic interstellar turbulence (\citealt{Chandran:2000}; \citealt{YanLazarian:2004}; \citealt{YanLazarian:2008}). However, detailed local observations of CR spectra (\citealt{Stone:2013}; \citealt{Adriani:2013}; \citealt{Aguilar:2015}; \citealt{Aguilar:2016}) seem to be inconsistent with these traditional transport theories (\citealt{Kempski:2022}; \citealt{Hopkins:2022}). Possible remedies include that interactions with turbulent magnetic mirrors play an important role (\citealt{LazarianXu:2021}; \citealt{Xu:2021}; \citealt{ZhangXu:2023}) or that the properties of the multiphase ISM need to be captured properly to model CR transport correctly (Armillotta et al., in prep.). Recently, a new class of propagation models has emerged, in which CR transport is instead regulated by intermittent strong scattering by rare large-amplitude magnetic-field fluctuations, in particular kinks and/or field reversals (\citealt{Lemoine:2023}; \citealt{Kempski:2023}; \citealt{Fielding:2023}; \citealt{Butsky:2024}). In this work, we focus on the intermittent scattering models and explore their possible connections to radio extreme scattering events (ESEs). 

\begin{figure}
    \includegraphics[width=\columnwidth]{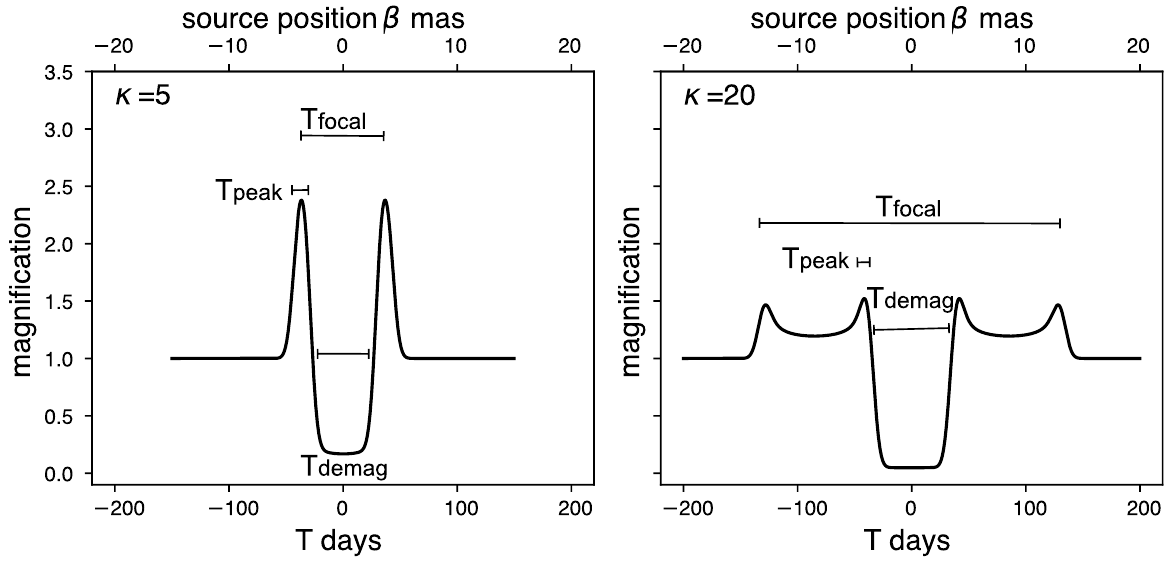}  
    \caption{A schematic lightcurve of a typical ESE. The two panels show examples of a source with an angular diameter of 1 mas and angular velocity 0.1 mas/day passing through overdense lenses with convergence values of $\kappa=5$ and $\kappa=20$ (equations~\ref{eq:kappa} and~\ref{eq:kappanumber}), respectively.}
    \label{fig:esetemperal}
\end{figure}

Radio waves are the most sensitive probes of small-scale structures in the warm ionized ISM with temperatures of order $10^4$ K. Intermittent structures have independently been proposed to explain various radio-wave scattering phenomena, including extreme scattering events (ESEs): dramatic week to year-long fluctuations in the radio flux of compact radio sources across multiple frequencies as shown by the schematic light curve in Figure~\ref{fig:esetemperal} \citep{1987Fiedler, 1993Cognard,2000Lazio,2001Lazio,2002Cim,2003Maitia,2008Senkbeil,2015Coles,2016Bannister,2018Kerr,2022Walker}. 
These flux modulations are best explained by a refractive lens in the ISM with a transverse scale of order an AU \citep{Romani:1987,1998Clegg}. In a few cases, lensed images have been observed using Very Long Baseline Interferometry (VLBI) due to its extremely high angular resolution \citep{2013Pushkarev,2023Koryukova}.
The ESE lightcurves, and more directly, the offset of the lensed images, imply an extremely large transverse gradient of the electron column density $N_e$,  with $dN_e / dx\sim \Delta n_e \ell /s   \sim 10^{3}-10^{5} \  {\rm cm}^{-3}$, where $\Delta n_e$ is the electron density variation across the lens, and $\ell$ and $s$ are the spatial scales of the lens along and perpendicular to the  line of sight (LOS), respectively. For a quasi-spherical plasma lens, i.e. $\ell \sim s$, this implies ionized number densities at least five orders of magnitude larger than the average density in the warm ionized ISM ($\sim$0.03 cm$^{-3}$). Such high-density plasma would, however, be drastically overpressurized relative to the ambient ISM and quickly disperse. This is known as the overpressure problem \citep{Stanimirovic:2018}. 
To address this issue, highly elongated structures with $\ell\gg s$ have been considered. The structure must be extremely straight, with a curvature radius  $r_c\gtrsim\ell^2/8s$, so that it does not bend away from the almost straight trajectory of radio waves. In this case, the large ratio of $\ell/s$, instead of the large $\Delta n_e$, is responsible for the large column density gradient. 

For a filament with three axes $s_1,s_2\ll \ell$, the chance of aligning the long axis with the LOS is so small that scattering will be dominated by unaligned filaments. Alternatively, thin flat current sheets with spatial scales $\ell_1,\ell_2\gg s$ have been proposed \citep{Romani:1987,1998Lestrade,2012PenKing, Jow2024}. A large electron column density gradient is naturally introduced by the small fraction of sheets that happen to align with the LOS. 
The sheet model has also been suggested to explain the strong scattering near the Galactic center \citep{2006GoldreichSridhar}, where standard Kolmogorov models of electron density fluctuations encounter difficulties in explaining scattering observations without becoming inconsistent with free-free radio emission data. 
At lower frequencies, cumulative studies of pulsar scintillation offer further evidence for the existence of density sheets. Pulsars scintillate due to multi-path scattering in the ISM. Detailed studies of nearby pulsars reveal that scattering often occurs within one or a few thin screens \citep{Stinebring:2001, Walker:2004, Cordes:2006}. Moreover, in some pulsars, scattering inside the screen is strongly inhomogeneous and occurs in localized clumps. \citealt{2010Brisken} used VLBI data to construct a scintillation-derived scattering image of PSR B0834+06, showing discrete scattering images aligned along a thin line on the sky plane. This highly localized and anisotropic scattering is difficult to reconcile with volume-filling, small-amplitude turbulent fluctuations. Consequently, refractive lenses formed by current sheets have been invoked to explain these features \citep{2014PenLevin, Simard:2018}. The connection between ESEs and pulsar scintillations is the subject of numerous studies \citep{2023Zhu,Jow2024}; notably, the current sheets proposed for ESEs, Galactic center scattering, and pulsar scintillations differ by orders of magnitude in thickness. Statistical studies of these events could provide new insights into the statistics of rare structures on different spatial scales in magnetized interstellar turbulence.

The plasma lenses responsible for ESEs have a transverse size of around an AU (\citealt{1987Fiedler}), which is intriguingly similar to the gyroradii of the energetically important GeV CRs.  
As argued by \citealt{2006GoldreichSridhar}, the density sheets are kept in pressure balance by an enveloping magnetic field that reverses direction. Such field reversals on $\sim$AU scales may be efficient scatterers of CRs (\citealt{Kempski:2023}; \citealt{Lemoine:2023}). Thus, it appears plausible that the same intermittent density/magnetic structures may be responsible for both strong radio-wave scattering and CR scattering. Exploring the possible connection between these two seemingly distinct astrophysical events is the primary objective of this paper. While we speculate that the scattering sheets are produced on small scales in interstellar turbulence, a detailed study of the origin of sheets capable of producing ESEs is beyond the scope of this paper. 

The paper is structured as follows. We summarize current observational constraints on CR scattering and ESEs in Section~\ref{sec:obs_constraints}. We introduce and motivate the geometry of magnetic/density structures considered in this work in Section~\ref{sec:geometry}. We calculate predicted rates of CR scattering and ESEs due to intermittent sheets in Section~\ref{sec:sheets} and compute parameter values for which the two scattering models can simultaneously be reconciled with existing observations. We discuss observational predictions in Section~\ref{sec:obs_predictions} and the validity of our model in Section~\ref{sec:discussion}, and summarize our key results in Section~\ref{sec:conclusions}.    

\section{Current Observational constraints } 
\label{sec:obs_constraints}
In this section, we summarize current observational constraints on CR transport, and the occurrence fraction and measured properties of ESEs. These observational constraints, alongside a list of parameter definitions used throughout this work, are summarized in Table~\ref{table:params}.

\begin{table*}
\centering
\caption{Parameters used throughout this work and their obervationally constrained values}
\label{table:params}
\begin{tabular}{p{1in}p{2.75in} p{2.75in}}
\hline
\hline
\textbf{Parameter} &  \textbf{Definition} & \textbf{Additional Notes / Observational Constraints}  \\ 
\hline
$\lambda_{\rm CR}$ &  Mean free path of $\sim$GeV cosmic rays  & Observed $\sim 1-10$ pc (\citealt{Zweibel:2013})  \\
$\lambda_\mr{PL}$ & Radio-wave mean free path due to aligned sheets  \\
$W_{\sr}$ &  ESE occurrence fraction & $\sim 10^{-3} - 10^{-2}$ for quasars, observed $\sim 0.007$ at 2.7 GHz in \citealt{1987Fiedler}  \\
$W_{\rm lens}$ &  The occurrence fraction of aligned sheets along the line of sight 
\\
 $\Omega_\mr{focal}$ & The angular area in the source plane affected by the lens\\
 $\Omega_\mr{lens}$& The angular area of the lens\\
$\kappa$ & $\Omega_\mr{focal}/\Omega_\mr{lens}$, the ratio of the angular area affected by the lens in the source plane to the angular size of the lens\\
$f_V$ &  Volume-filling fraction of sheets
\\
$s$ &  Sheet short axis &  Of order $\sim 0.1-10$ AU for ESEs  \\
$\ell$ &  Sheet long axis   \\
$D_\mr{lens}$ & Distance between the lens and observer \\
$D_\mr{src}$ & Distance between the source and observer \\
$D_\mr{ls}$ &Distance between the source and lens \\
$\Delta n_e$ &   Electron density variation across sheet & Density variance $\sim 0.01-0.1 \ {\rm cm^{-3}}$ in warm ISM
\\
$dN_e/dx$ &  Gradient of the line-of-sight electron column density $N_e$ in the sky plane & $\gtrsim 1000 \ {\rm cm^{-3}}$ to explain ESE
\\
$\beta_\mr{p}$ & Ratio of thermal gas pressure to magnetic pressure\\
$\beta$ & Angular position of the source\\
$\theta$ & Angular position of the lensed images \\
$\hat\alpha$ &  Refractive deflection angle of radio waves&$\hat\alpha=-(r_e\lambda^2/2\pi)dN_e/dx$\\
$\alpha$ & Reduced deflection angle &$\alpha=\hat\alpha(D_{\rm ls}/D_{\rm src})$\\\
$\nu$ & Observation frequency \\
$\lambda$ & Observation wavelength \\

\hline
Other Observables \\
\hline
$\Delta$DM & Dispersion measure: $\int \Delta n_e dl$&$10^{-3}-10^{-1}$pc cm$^{-3}$ measured from pulsar ESEs \citep{2015Coles,2018Kerr}\\
$\Delta$RM & Rotation measure: $\int \Delta n_e B_\parallel dl$& $<10$ rad~m$^{-2}$\citep{1996Clegg}\\
\hline
\hline
\end{tabular}
\end{table*}

\subsection{ESE constraints}
In this subsection, we summarize the current observational constraints on ESEs. The statistics remain highly uncertain, because of difficulties in distinguishing ESEs from intrinsic flux modulations of quasars. This challenge arises because only a small subset of observations includes multiple frequencies, VLBI measurements, or changes in electron column density measured from dispersion measure (DM).
\begin{itemize}
			\item The classic Fiedler ESE events \citep{1987Fiedler} exhibit a double-peak structure in the 2.7~GHz light curves, with the peaks separated by approximately 80 days (labelled as $T_\mr{focal}$ in Figure~\ref{fig:esetemperal}) and a demagnification phase of approximately 50 days  (labelled as $T_\mr{demag}$ in Figure~\ref{fig:esetemperal}). Similar lightcurves have been seen with other quasars and pulsars with duration ranging from weeks to years. 
	\item \cite{1987Fiedler, 1994Fiedler} estimated the occurrence fraction of ESEs, i.e. the fraction of sources undergoing ESEs at any given time, from quasars at 2.7~GHz to be $\sim 0.01$. 
    \item For a small number of quasars, plasma-lensing images aligned in a linear configuration with highly chromatic offsets have been observed using VLBI during ESEs \citep{2013Pushkarev,2023Koryukova}. 
    \item For a small number of pulsars with observed ESEs, the measured DM variation is around $10^{-3}-10^{-1}$ pc cm$^{-3}$ \citep{2015Coles,2018Kerr}, with the change of rotation measure RM$=\int \Delta n_e B_\parallel dl<10$ rad~m$^{-2}$ \citep{1996Clegg}. 
\end{itemize} 
These observational constraints are also summarized in Table~\ref{table:params}, and 
we will derive physical constraints on the plasma lensing from the observables in section~\ref{sec:rate_heuristic} and~\ref{sec:sheet_params}.

\begin{figure*}
  \centering
\includegraphics[width=0.99\textwidth]{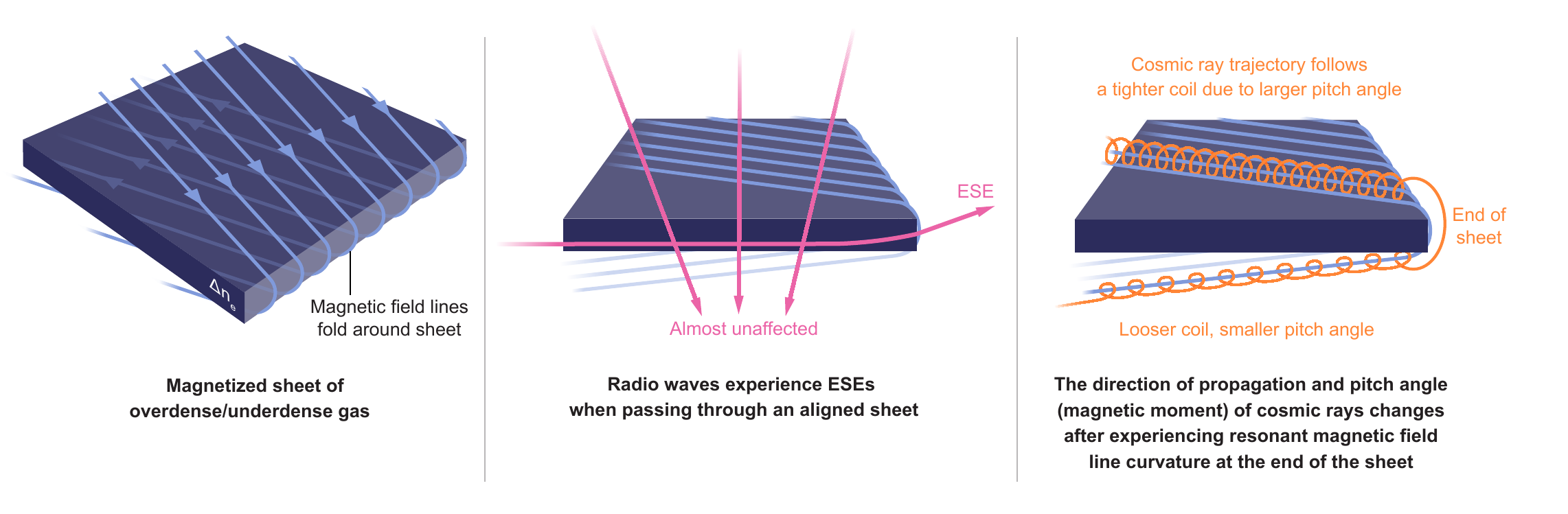}
\caption{Schematic of current/density sheets considered in this work. Left: overdense current sheet with density perturbation $\Delta n_e$ kept in pressure balance by a folded magnetic field. For layers of multiple current sheets, as often seen in simulations of large-amplitude MHD turbulence, there are both overdense and underdense sheets. Middle: significant radio wave scattering, i.e., an ESE, occurs when the sheet is aligned favourably for the light ray to traverse a large column density parallel to the sheet. Right: the folded field lines are responsible for strong CR scattering, which occurs at the end of the sheet when particles experience resonant field-line curvature. We discuss alternative sheet geometries in Appendix~\ref{sec:alternative_sheet}.  \label{fig:scattering_schematic}}
\end{figure*}

\subsection{Constraints on CR scattering}
Measurements of secondary CRs, such as boron or beryllium, produced in the ISM through spallation reactions of primary CRs accelerated in supernovae, suggest an energy-dependent mean free path in the Galaxy that follows the approximate scaling (\citealt{Amato:2018}),
\begin{equation}\label{eq:CR_mfp_MW}
    \lambda_{\rm CR} \sim 1 \ {\rm pc} \ \Big( \frac{E}{\rm GeV} \Big)^{0.5}.
\end{equation}
We note that there is some ambiguity as to what the GeV CR mean free path is in the Galaxy, with values in the literature ranging from 0.1 pc to 10 pc (\citealt{Zweibel:2013}; \citealt{Amato:2018}; \citealt{Evoli2020}; \citealt{DeLaTorreLuqueL2021}; \citealt{Butsky:2024}). This uncertainty comes from uncertain spallation reaction cross sections and from the fact that one typically needs to choose a  `CR halo size' when modelling the escape of CRs from the Galaxy. Mean free path values $\lesssim$ pc are favoured by models that consider small halo sizes $\lesssim$ a few kpc, while values $\gtrsim 1$ pc are favoured by models that account for the confinement of CRs in the much larger circum-galactic medium (CGM).  We favor the latter for reasons discussed in \citealt{Kempski:2022} and \citealt{Hopkins:2022}.

\section{Geometry of magnetized density sheets }\label{sec:geometry}
In this section, we describe a physically motivated model of magnetized density sheets that can simultaneously contribute to both CR scattering and ESEs. We discuss an alternative sheet model in Appendix~\ref{sec:alternative_sheet} and show why quasi-spherical structures are unlikely candidates in Appendix~\ref{sec:spherical_geometry}.

Very straight and large-aspect-ratio, aligned density sheets with thicknesses $\sim$AU have been proposed as a possible explanation for observed ESEs and pulsar scintillation (\citealt{2006GoldreichSridhar}; \citealt{2014PenLevin}; \citealt{Simard:2018}; \citealt{Jow2024}). In particular, if sufficiently aligned and elongated, density sheets do not require extreme density fluctuations to produce large gradients in electron column density. More recently, current sheets associated with folded magnetic-field structures have been proposed as a source of CR scattering (\citealt{Kempski:2023}), with current sheets of thicknesses $\gtrsim$AU responsible for the scattering of energetically important $\sim$GeV CRs. This raises the important question: are CRs and radio waves strongly scattered by the same sheet-like structures in the ISM? A key requirement for this unified picture is that the folded magnetic fields responsible for CR scattering also correspond to sheets of overdense or underdense plasma. This is indeed the case if the sheets are isothermal and in pressure balance,\footnote{Even if the sheets are produced by a turbulent flow, ram pressure from the velocity field around the sheet can nevertheless be ignored in the transverse pressure balance because the velocity difference in the direction of the short axis of the sheet is highly sub-Alfv\'enic, see discussion in Section~\ref{sec:tearing}.}
\begin{equation} \label{eq:sheet_p_bal}
    \frac{\delta \rho}{\rho} \sim \beta_\mr{p}^{-1} \frac{\delta B}{B} \sim \beta_\mr{p}^{-1},
\end{equation}
where $\delta \rho/\rho$ is the fractional density perturbation in the sheet,  $\beta_\mr{p}=8\pi p_g / B^2$ is the ratio of thermal gas to magnetic pressure evaluated outside the sheet, and in the last step we used that in a reversing magnetic-field structure $\delta B/B \sim 1$. Eq.~\eqref{eq:sheet_p_bal} is a good approximation for $\beta_\mr{p} \gtrsim 1$, which we assume to hold in a large fraction of the volume in the warm ISM. To maintain pressure balance under such conditions, current sheets therefore host overdense plasma where the field lines reverse direction, which may act as a strong plasma lens (\citealt{2006GoldreichSridhar}). Thus, in this work we shall mostly focus on current/density sheets as illustrated by the schematic in Figure~\ref{fig:scattering_schematic}, although we also discuss alternative sheet models in Appendix~\ref{sec:alternative_sheet}. We assume density sheets kept in pressure balance by folded magnetic field lines, as shown in the left panel of Figure~\ref{fig:scattering_schematic}. Radio waves are scattered significantly by such sheets if their direction of propagation is aligned with the long axis of the sheet (middle panel). Scattering of CRs occurs at the location where the magnetic field folds onto itself and the particle experiences resonant field-line curvature, i.e., if the field lines bend on a scale of order the CR gyration radius $s \sim c / \Omega$ (\citealt{Lemoine:2023}; \citealt{Kempski:2023}), where $c$ is the speed of light and $\Omega$ is the relativistic gyrofrequency. 

While not all current sheets are necessarily associated with magnetic folds and sharply bent field lines that can scatter CRs, throughout this work we assume that at least an order unity fraction of strong current sheets (i.e., ones with reversing magnetic-field lines and therefore large density perturbation) are due to folded field lines. This is motivated by the theory of how field reversals are formed in the small-scale turbulent dynamo (\citealt{Schekochihin:2004}; \citealt{Rincon:2019}), which is plausibly qualitatively similar in MHD turbulence with weak guide field \citep{Kempski:2023}. We note that how the plasma maintains pressure balance may be different from equation~\eqref{eq:sheet_p_bal} depending on the importance of cooling (\citealt{Fielding:2023}). However, proper treatment of multiphase effects is beyond the scope of this work. While the density sheet in Figure~\ref{fig:scattering_schematic} represents an overdense region, if there are layers of multiple current sheets as often seen in simulations of large-amplitude MHD turbulence (\citealt{Kempski:2023}), both overdense and underdense sheets are expected to be present as the magnetic-field strength goes through a series of maxima and minima.

\begin{figure}
  \centering
    \includegraphics[width=\columnwidth]{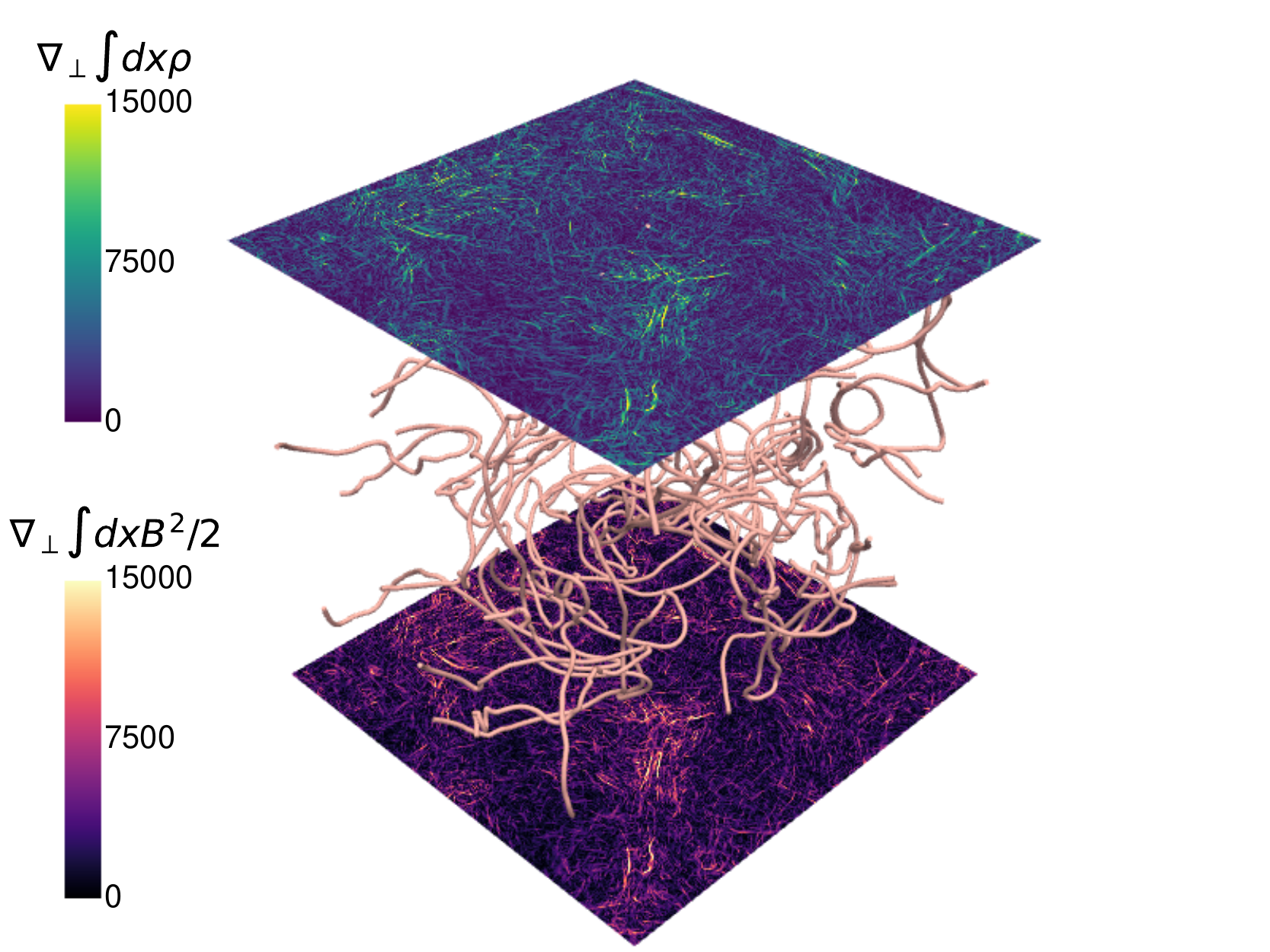}  
    \caption{Slices of in-plane gradients of column density and integrated magnetic pressure (arbitrary units) from a $1024^3$ subsonically driven simulation of MHD turbulence with weak guide magnetic field ($\delta B/B_0 \approx 4$) taken from \citet{Kempski:2023}. The most intense density sheets, which produce strong radio-wave scattering, are also sheets of strongly varying magnetic-field strength, qualitatively consistent with the picture of pressure-balanced current sheets (eq.~\ref{eq:sheet_p_bal} and Figure~\ref{fig:scattering_schematic}). Between the two slices, we show the 3D morphology of the turbulence using a sample of highly-tangled magnetic field lines, characterized by numerous kinks and reversals that scatter CRs (\citealt{Lemoine:2023}; \citealt{Kempski:2023}).}
 \label{fig:dBB4_slices}
\end{figure}

In Figure~\ref{fig:dBB4_slices}, we demonstrate how equation~\eqref{eq:sheet_p_bal} is realized in a simulation of driven isothermal MHD turbulence taken from \cite{Kempski:2023}, which was run at resolution $1024^3$ in a regime with weak guide magnetic field ($\delta B/B_0 \approx 4$) and sonic Mach number $\approx 0.5$. The 2D slices show the magnitude of the in-plane gradient of the projected (ionized) gas column density and magnetic pressure integrated along the same line of sight, demonstrating that the most intense density sheets are also sheets of strongly varying magnetic-field strength. Between the two slices, we show the 3D morphology of the turbulence using a sample of highly-tangled magnetic field lines, characterized by numerous kinks and reversals.

Finally, we note that while current sheets are generally characterized by three length-scales, $\ell_1, \ell_2$ and $s$, with $\ell_1$ being the longest axis, $\ell_2$ the intermediate axis and $s$ being the shortest axis,\footnote{The largest dimension $\ell_1$ typically corresponds to the coherence length of the magnetic field along itself, $\ell_2$ represents the width of the sheet in the direction of the current density, $\mathbf{j}=c/(4 \pi) \mathbf{\nabla \times B}$, and $s$ is the current sheet thickness in the direction of field reversal.} in this work we restrict ourselves to the simplified case of perfect sheet-like geometries with $\ell_1 \sim \ell_2 \sim \ell $. While this is often assumed in the ESE/pulsar scintillation literature, we note that this assumption is not perfectly realized in simulations of MHD turbulence (\citealt{Zhdankin:2013}; \citealt{Galishnikova:2022}).

\section{Scattering by sheets}\label{sec:sheets}
We now consider the scattering of both CRs and radio waves by the magnetized density sheets illustrated in Figure~\ref{fig:scattering_schematic}. Motivated by observations of ESEs, we consider the sheets to be straight. We briefly discuss physical processes that may be important for setting the geometry of sheets in Section~\ref{sec:dissipation}, but a detailed study of the origin of sufficiently straight sheets is beyond the scope of this work. Here, the straightness of the sheets is an observationally motivated ingredient in our model. 

\subsection{Cosmic ray scattering in folded current/density sheets}

As illustrated in Figure~\ref{fig:scattering_schematic}, in a folded magnetic-field structure with length $\ell$ and thickness $s$, CRs propagate the parallel length of the sheet $\ell$ and get scattered at the end of the fold if they experience resonant field-line curvature, i.e., if their local gyroradius is comparable to the length-scale over which the field line bends \citep{Lemoine:2023, Kempski:2023}. For typical values of the Galactic magnetic field (of order a few $\mu$G), the gyroradius of a GeV CR is of order $0.1$ AU. However, magnetic-field strengths are significantly reduced in regions where the field lines bend sharply (\citealt{Schekochihin:2004}; \citealt{Rincon:2019}).\footnote{We note, however, that the dependence of magnetic-field strength on field-line curvature may be weaker in supersonic turbulence (\citealt{Kriel:2023}).} As a result, the local CR gyroradius is significantly increased \citep{Lemoine:2023, Kempski:2023} and may instead become of order $10$ AU. This is illustrated by the increased radius of curvature of the orange CR trajectory in the right panel of Figure~\ref{fig:scattering_schematic}. If ESEs are indeed due to sheets, observations constrain their thicknesses to be of order $\sim 1$ AU, which is in the range of gyroradii of $\sim$GeV CRs in such a magnetic fold. The CR mean free path due to scattering by $s \sim \ {\rm AU}$ sheets is
\begin{equation}\label{eq:cr_mfp_fold}
    \lambda_{\rm CR} \sim \frac{\ell}{f_V (s \sim  {\rm AU})},
\end{equation}
where $f_V$ is the volume filling fraction of the sheets with dimensions ($\ell$,$\ell$,$s$). Note that in this picture of CR scattering, the CR mean free path measured in the Galaxy~\eqref{eq:CR_mfp_MW} directly relates the required volume-filling fraction to the parallel lengths of the sheets. We will use this relation to test whether the measured CR mean free path, if indeed due to scattering by elongated magnetic structures, is consistent with occurrence fraction of ESEs. 

\begin{figure*}
  \centering
  \begin{minipage}{0.28\linewidth}
\includegraphics[width=\columnwidth]{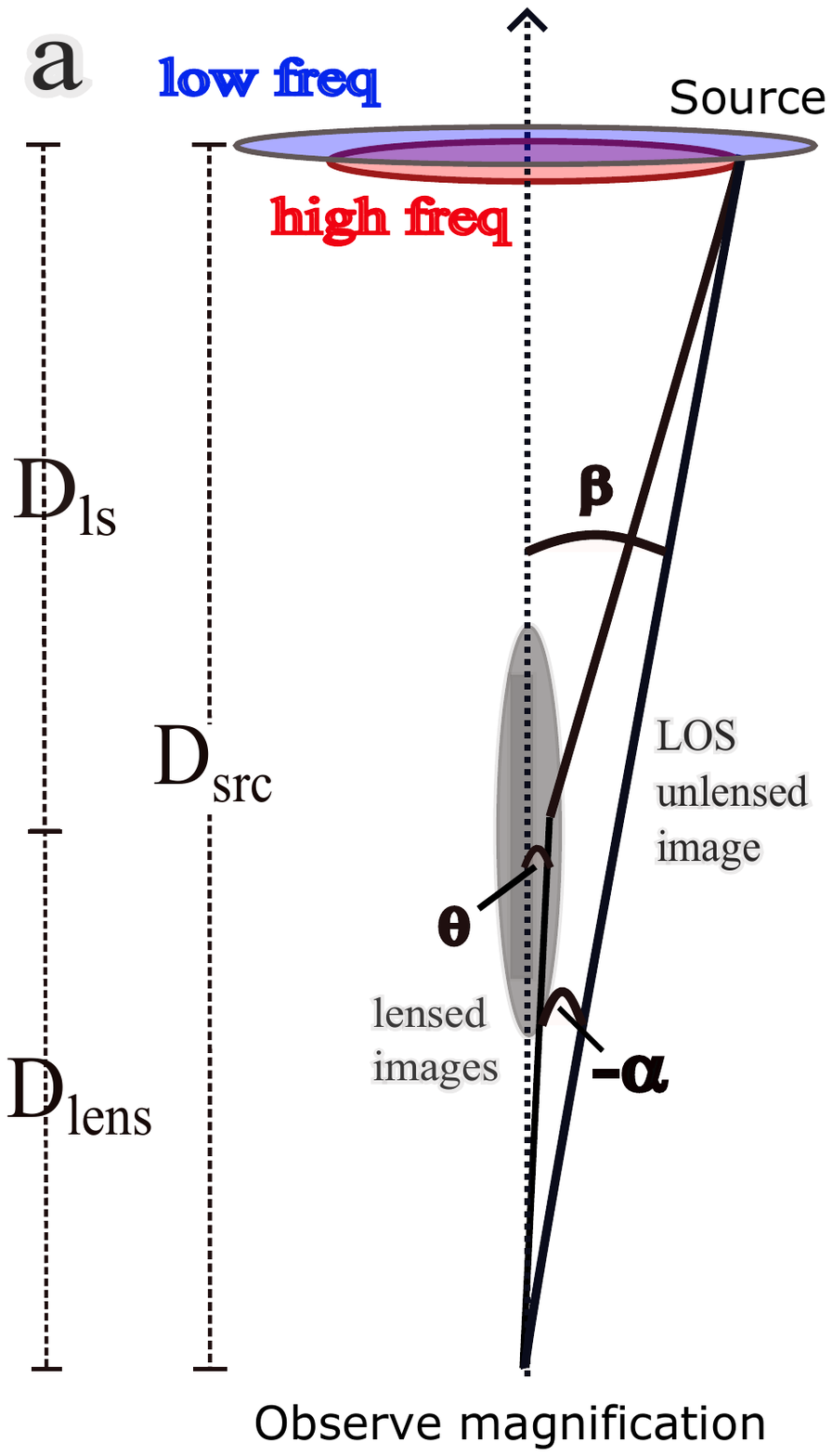}      
  \end{minipage}
    \begin{minipage}{0.28\linewidth}
    \includegraphics[width=\columnwidth]{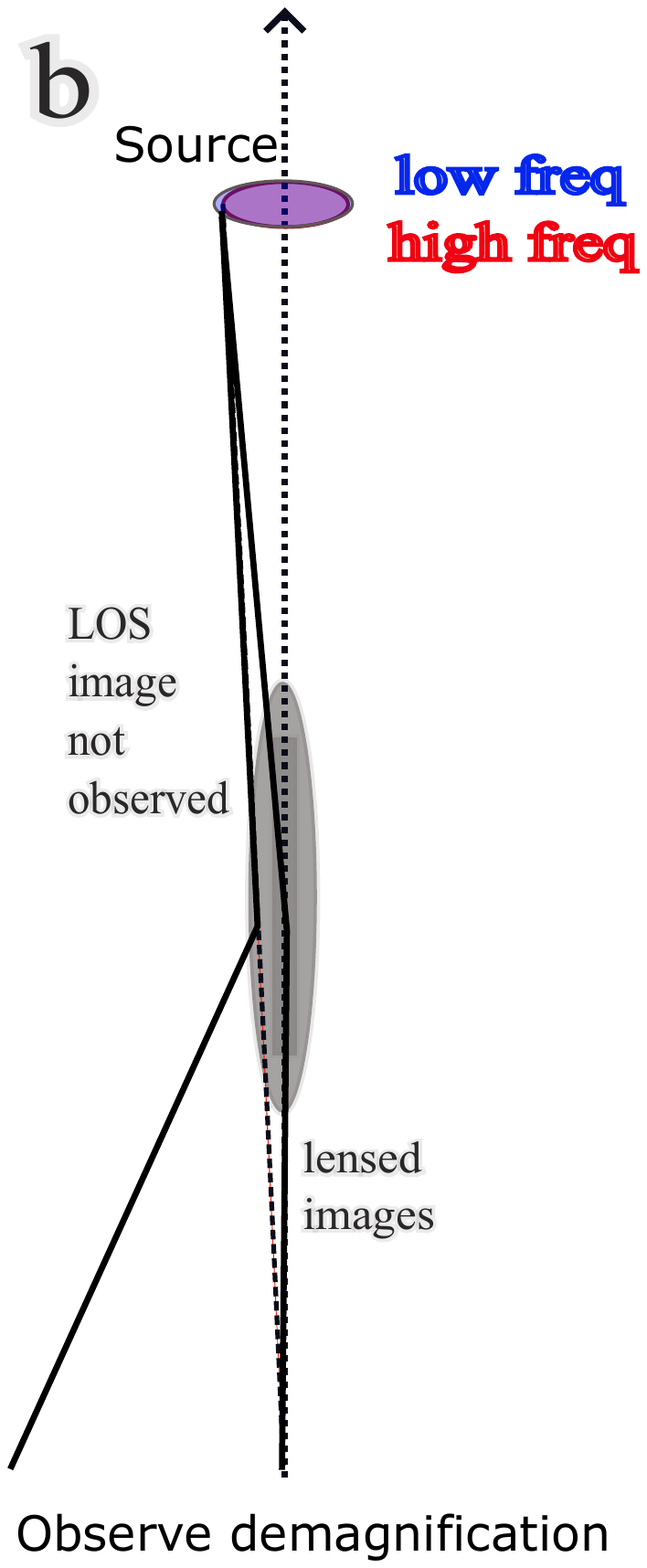}      
  \end{minipage}
  \begin{minipage}{0.42\linewidth}
    \includegraphics[width=\columnwidth]{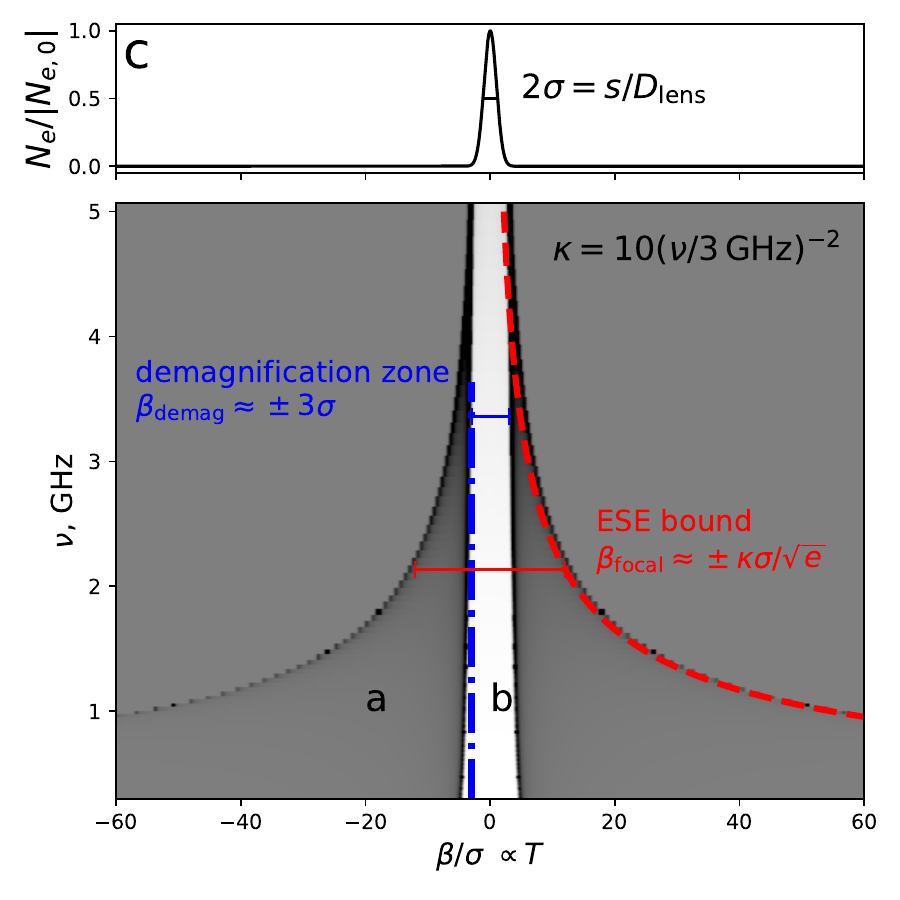}      
  \end{minipage}
    \caption{The lensing behavior of an aligned sheet. (a) When the source is close to the direction of the sheet, the observer can receive the light bent by the sheet, as well as unlensed light from the LOS, and hence observe magnification of the source. This regime is labelled ``a'' in the rightmost panel. The dashed line with arrow is the fiducial fixed sky direction, relative to which the source angular position $\beta$ is defined. The largest bending angle introduced by the sheet sets the size of the angular region on the source plane affected by the lens, which can be used to estimate the occurrence fraction of ESEs. (b) 
    When the source is behind the aligned sheet, the LOS light is bent away and the observer only sees the faint lensed images and hence observes demagnification of the source. The demagnified region (labelled ``b'' in the rightmost panel) is determined by the thickness of the sheet and is almost achromatic. (c top) The angular profile of the column density projected on the sky can be approximated by a Gaussian with angular width $\sigma$ approximately equal to half its physical width along the sheet normal, divided by the distance to the sheet. (c bottom) The magnification (background color, darker means higher magnification) is a function of angular source position $\beta$. The source will be lensed in the region between $\pm\beta_\mr{focal}$ (Eq.~\ref{eq:beta_focal} and~\ref{eq:betaout}) and will be demagnified between $\pm\beta_\mr{demag}$ (Eq.~\ref{eq:beta_demag} and~\ref{eq:betademag}). The red and blue dashed lines are analytical approximations of $\pm\beta_\mr{focal}$ and $\pm\beta_\mr{demag}$, which are used to estimate ESE occurrence fraction and sheet thicknesses, respectively.    \label{fig:lensFreq} }
\end{figure*}

\subsection{ESEs due to current/density sheets}
A radio wave will be strongly scattered by a density sheet when it is sufficiently aligned with its direction of propagation (Figure~\ref{fig:scattering_schematic}). We calculate the probability of alignment and corresponding ``ESE mean free path'' in Section~\ref{sec:chance_align} below. We then compute the occurrence fraction of sources undergoing ESEs due to aligned straight sheets in Section~\ref{sec:rate_heuristic}, which can be directly compared to observed rates (e.g., \citealt{1987Fiedler}; see Table~\ref{table:params}). 

\subsubsection{Probability of sheet alignment} \label{sec:chance_align}
We denote $\lambda_{\rm PL}$ as the typical distance (or mean free path) that a radio wave propagates before it encounters an aligned plasma lens, which causes an ESE. For straight sheets with dimensions ($\ell$,$\ell$,$s$), $\lambda_{\rm PL}$ is set by the number density $n_s$ of such structures, the probability of alignment $P_{\rm align} \sim s/\ell$, and the aligned cross section $A \sim s \ell$:
\begin{equation}  
    \lambda_{\rm PL} \sim (n_s A P_{\rm align})^{-1} \sim  \frac{\ell}{f_V} \frac{\ell}{s}.
\end{equation}
In the last step we related the number density to the volume-filling fraction, $n_s \sim f_V / (\ell^2 s)$. The mean free path for strong scintillation or an ESE is therefore larger than the CR mean free path by the inverse of the alignment probability $s / \ell$, giving
\begin{equation}\label{eq:lambda_pl_cr}
    \lambda_{\rm PL} \sim \lambda_{\rm CR} \frac{\ell}{s}.
\end{equation}
This comes from the fact that, while CRs are constrained to propagate along the magnetic field along the sheet (and so alignment is guaranteed), this is not true for radio waves. Assuming randomly oriented sheets over a distance $L$, the probability of having an aligned structure $W_{\rm lens}$ along a specific line of sight is,
\begin{equation}
    W_{\rm lens} \sim \frac{L f_V s}{\ell^2}.
    \label{eq:wl}
\end{equation}
The aligned structure corresponds to a column density gradient,
\begin{equation}
    \frac{dN_e}{dx} \sim \Delta n_e \frac{ \ell}{s}.
\end{equation}
We note that $W_{\rm lens}$ is generally not equal to the occurrence fraction of strong scattering events because, in the presence of finite lensing, the observer, lens, and source do not need to be perfectly aligned. We show this based on simple arguments in Section~\ref{sec:rate_heuristic} and using a more rigorous lensing calculation in Appendix~\ref{sec:gaussian_lens}. 

\subsubsection{ESE rates from aligned sheets} \label{sec:rate_heuristic}
An aligned density sheet acts as a plasma lens for background sources. 
We define the focal region of the lens as the area on the source plane where the lens has sufficient bending power to refract light onto the observer. 
This focal region is analogous to the field of view of a telescope, which also functions as a lens. 
The focal region of this plasma lens is frequency dependent due to the strongly chromatic nature of the refractive index. 
The occurrence fraction of ESEs for a background point source $W_\mr{src}$ is proportional to the total angular size of the focal area $\Omega_\mr{focal}$ at a given frequency from all the aligned sheets:
\begin{align}
W_\mr{src} = W_\mr{lens}\, \frac{\Omega_\mr{focal}}{\Omega_\mr{lens}} := W_\mr{lens} \kappa,
\label{eq:wsrc2lens}
\end{align}
where $W_\ls$ is again the chance of encountering an aligned sheet at a given LOS; $\Omega_\ls=ls/D_\mr{lens}^2$ is the 
projected angular area of the lens, where $D_\mr{lens}$ is the distance to the lens. The convergence $\kappa \equiv \Omega_\mr{focal} / \Omega_\mr{lens}$ introduced in the last step quantifies the strength of the plasma lens. Thus, at fixed $W_\mr{lens}$ the occurrence fraction of ESEs is higher when the sheets produce stronger lensing.  

When the lensing effect is weak, light from a background source observed at Earth experiences the sheet only when the source is right behind the sheet, and hence $\Omega_\mr{focal}\sim\Omega_\mr{lens}$. When the lensing effect is strong, as in the case of ESEs,  light can be significantly bent by the aligned sheets, as shown in Fig.~\ref{fig:lensFreq}. As a result, the sheet position does not have to be exactly along the line connecting the observer and the source for ESE/scintillation to occur.  
In this case, the angular size of the focal region $\Omega_\mr{focal}$ is related to the reduced deflection angle of the lens, $\alpha$ (see Fig.~\ref{fig:lensFreq}), which is set by the gradient of the light path difference: 
\begin{equation} 
    \alpha=\frac{D_\mr{ls}}{D_\mr{src}}\nabla_x \int n dz ,
    \label{eq:alpha}
\end{equation}
where $n$ is the plasma refractive index, $x$ is an axis in the sky plane, $z$ is the axis along the line of sight, $D_\mr{ls}$ is the distance between the lens and the source, and $D_\mr{src}$ is the distance between the Earth and the source. Variations in the refractive index $\Delta n$ are caused by variations in electron number density,
\begin{equation}
    \Delta n = \frac{r_e c^2}{2 \pi \nu^2} \Delta n_e \ \sim 4 \times 10^{-12} \ \frac{\Delta n_e}{0.1 \ {\rm cm}^{-3}} \Big(\frac{1 {\rm GHz}}{\nu}\Big)^2,
\end{equation}
where $r_e$ is the classical electron radius, $\nu$ is the radio-wave frequency and $\Delta n_e$ is the electron density change (e.g., in the sheet as in eq.~\ref{eq:sheet_p_bal}). 

The reduced deflection angle $\alpha$ is only significant along the short axis of the sheet along which the electron column density has a large gradient. Therefore, for an aligned sheet in equation~\ref{eq:alpha}, the integration length-scale along the $z$ axis is of order $\ell$, while  the electron column density changes across half the thickness of the sheet, hence $\nabla_x\sim 2/s$ and the bending angle can be expressed by: 
\begin{align}
    \alpha_s \approx \Delta n \frac{2\ell}{s}.
\end{align} 
Here we have used $D_\mr{ls}/D_\mr{src}\approx1$ for quasars\footnote{For the scattering of pulsar light, both the lens and source are in the Milky Way. Therefore, we have to replace the $D_\mr{lens}$ in equation~\eqref{eq:kappanumber} and~\eqref{eq:wsrcnumber} with $D_\mr{eff}=D_{ls}D_\mr{lens}/D_\mr{src}$. The $D_\mr{eff}$ is also usually of the order of 100pc at high galactic latitudes to kpc at lower latitudes. } as the lens is in the Milky Way and hence much closer than the source.  The factor of 2 comes from the assumption that the sheet is symmetric and so the gradient is set by the half-thickness of the sheet. 
The reduced deflection angle is negligible along the long axis of the sheet, with $\alpha_\ell \sim \Delta n \ll 5\times 10^{-9} (s/\mr{AU})/(D_\mr{lens}/\mr{kpc})\ll \ell/D_\mr{lens}$, in other words the deflection is much smaller than the angular size of the lens. 

A source at angular position $\beta$ is affected by the lens if there exists an $\alpha$ such that $\beta=\theta-\alpha$. The angles are defined as shown in Figure~\ref{fig:lensFreq} panel a. As the lensed image always come from the direction of the lens, when the deflection angle is large, $\alpha\gg\theta$ and $\beta\approx \alpha$. Hence, we can use the maximum deflection angle $\alpha_s$ to estimate the size of the focal area $\Delta \beta_\mr{focal}$. The convergence $\kappa$ can then be estimated with: 
\begin{align}\label{eq:kappa}
\kappa &= \frac{\Omega_\mr{focal}}{\Omega_\mr{lens}}\\
&\sim \mr{max}\left(1,\frac{2\alpha_s D_\mr{lens}}{s}\right) \sim \mr{max}\left(1,4D_\mr{lens} \Delta n \frac{\ell}{s^2}\right)  .\nonumber
\end{align}
The additional factor of 2 in the last step comes from the fact that the sheet can bend the image in both directions and hence, the total focal length is twice the individual bending angle.  We provide a more rigorous calculation of the convergence $\kappa$ in Appendix~\ref{sec:gaussian_lens}. For models involving asymmetric sheets, the factor of 2 may be dropped. 

For $\kappa>1$,  and hence $ \alpha_s > s/ D_\mr{lens}$,  we have: 
\begin{align}
\label{eq:kappanumber}
\kappa &= 4 \lambda^2 \frac{r_e}{2\pi} D_\mr{lens} \Delta n_e \frac{\ell} {s^2}\\
&= 30 \left(\frac{\nu}{\mr{GHz}}\right)^{-2}\left(\frac{D_\mr{lens}}{\mr{kpc}}\right)\left(\frac{\Delta n_{e}}{0.1\mr{cm}^{-3}}\right)\left(\frac{\ell}{10^4 \mr{AU}}\right)\left(\frac{s}{\mr{AU}}\right)^{-2}.\nonumber
\end{align}
The total duration of the ESEs is: 
\begin{align}
    T_\mr{focal} = \frac{\Delta\beta_\mr{focal}}{\omega_\perp} \approx \frac{\kappa s}{D_\mr{lens}\omega_\perp},
    \label{eq:Tfocal}
\end{align}
where $\omega_\perp=\omega_\beta\cos\phi$ is the relative angular velocity of the source along the sheet normal, and $\phi$ is defined as shown in Figure~\ref{fig:passAngle}. We will discuss how it is measured in Section~\ref{sec:sheet_params}. $\beta_\mr{focal}$ is the angle that corresponds to the edge of the angular focal region affected by the lens,
\begin{align}
\Delta\beta_\mr{focal} = C_1 \frac{\kappa s}{D_\mr{lens}},
    \label{eq:beta_focal}
\end{align}
where $C_1$ is an order unity constant that depends on the shape of the lens. For a lens with a  Gaussian density profile, $C_1 \approx 1/\sqrt
e$ (Appendix~\ref{sec:gaussian_lens}).  
 
Combining~\eqref{eq:wl} and~\eqref{eq:wsrc2lens}, we have the frequency-dependent occurrence fraction of ESEs: 
\begin{align}	\label{eq:wsrcnumber}
    W_\mr{src} &= W_\mr{lens}\, \frac{\Omega_\mr{focal}}{\Omega_\mr{lens}} 
    \sim \frac{L f_V s}{\ell^2} \kappa \\
	&\sim0.07 \left(\frac{\nu}{\mr{GHz}}\right)^{-2}\left(\frac{D_\mr{lens}L}{\mr{kpc}^2}\right)\left(\frac{\Delta n_{e}}{0.1\mr{cm}^{-3}}\right) \nonumber \\
	&\times \left(\frac{s}{\mr{AU}} \right)^{-1}\left(\frac{l}{\mr{10^{4}AU}}\right)^{-1} \left(\frac{f_V}{\mr{10^{-3}}}\right). \nonumber 
\end{align}
 Interestingly, in this model the occurrence fraction depends on $f_V$ and $\ell$ only through their ratio, which in the sheet model also corresponds to the CR mean free path~\eqref{eq:cr_mfp_fold}. This fact will be important when we assess whether the same sheets may simultaneously provide an explanation for observed rates of CR scattering and ESEs. 

\subsection{Are ESEs and CR scattering rates consistent?} \label{sec:consistency}
Assuming that the thin currents sheets are terminated by a sharp bend that can scatter CRs as in Figure~\ref{fig:scattering_schematic} (see discussion in Section~\ref{sec:geometry}), we can estimate the mean free path of $\sim$GeV CRs using the observed rates of ESEs and equations~\eqref{eq:cr_mfp_fold} and~\eqref{eq:wsrcnumber}: 
\begin{align}
&\lambda_\mr{CR}^\mr{pred.} \sim \frac{\ell}{f_V} \\
    & \sim 10\ \mr{pc} \left(\frac{W_\mr{src}}{0.01}\right)^{-1}\left(\frac{\nu}{3\mr{GHz}}\right)^{-2}
    \left(\frac{L\,D_\mr{lens}}{\mr{kpc}^2}\right)
    \left(\frac{\Delta n_e}{0.03\mr{cm}^{-3}}\right)
    \left(\frac{s}{\mr{AU}}\right)^{-1} . \nonumber
\end{align}
Therefore, for sheets with density fluctuations similar to the mean density of the warm ionized phase of the ISM ($\sim$0.03 cm$^{-3}$), the expected mean free path of CRs is roughly consistent with the observed ESE occurrence fraction.

\begin{figure}
  \centering
    \includegraphics[width=0.48\textwidth]{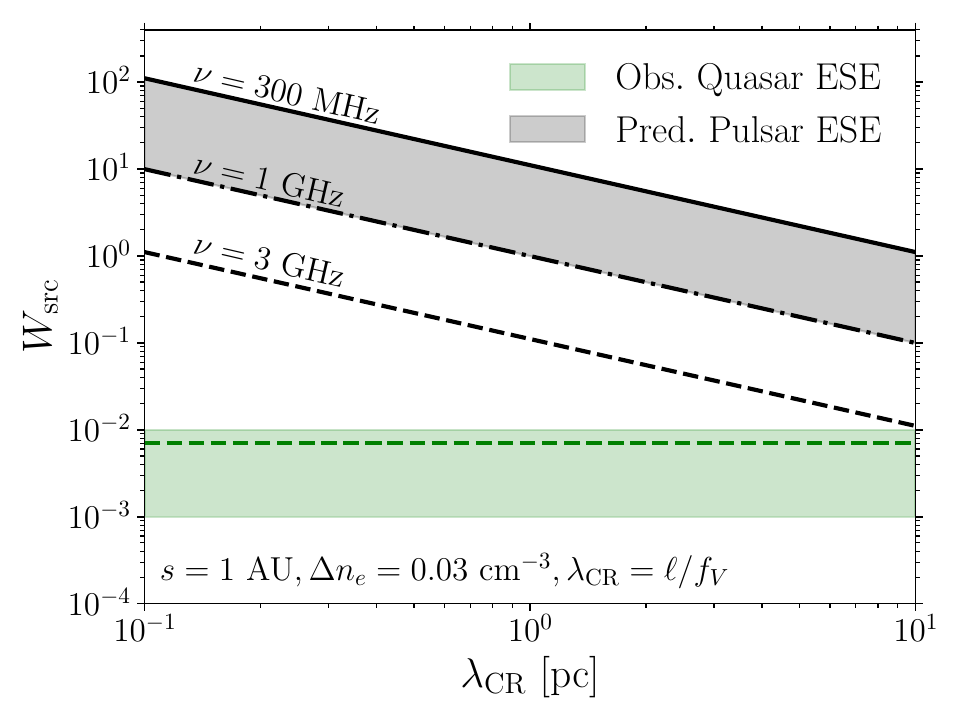}
    \caption{Occurrence fraction of radio sources being lensed by aligned plasma sheets of thickness ${\sim}1~{\rm AU}$ calculated from~\eqref{eq:wsrcnumber} in three different frequency bands as a function of the $\sim$GeV CR mean free path. We assume that the long axis of the sheets $\ell$ and volume filling factor $f_V$ are constrained by the CR mean free path, $\lambda_{\rm CR} \sim \ell / f_V$. For a fixed $\lambda_{\rm CR}$, $W_{\rm src}$ is then independent of $\ell$. The green shaded region shows rough occurrence fraction  of quasar ESEs, with the green dashed line representing the rate $\approx 0.007$ at 2.7~GHz reported in \citet{1987Fiedler,1994Fiedler}.  The grey shaded region shows the predicted approximate occurrence fraction of pulsar ESEs observed at lower frequencies as a function of $\lambda_{\rm CR}$. Assuming that CR scattering and ESEs are caused by the same density/magnetic sheets of thickness $\sim 1$AU and with $\Delta n_e \sim 0.03 \ {\rm cm}^{-3}$, a CR mean free path of order $10$ pc is consistent with observed occurrence fraction of quasar ESEs (observed at frequencies $>$~GHz).}
     \label{fig:Wsrc_lambdaCR}
\end{figure}

We summarize the relationship between CR scattering mean free paths and the ESE occurrence fraction in Figure~\ref{fig:Wsrc_lambdaCR}, which shows the occurrence fraction of radio sources being lensed by aligned plasma sheets of thickness $\sim 1$~AU  in three different frequency bands as a function of the $\sim$GeV CR mean free path. The occurrence fraction is calculated from~\eqref{eq:wsrcnumber} and assumes that the long axis of the sheets $\ell$ and volume filling factor $f_V$ are constrained by the CR mean free path, $\lambda_{\rm CR} \sim \ell / f_V$. The green shaded region shows the rough occurrence fraction $W_{\rm src}$ of quasar ESEs, with the green dashed line representing the fraction $\approx 0.007$ reported in \cite{1987Fiedler}.  The grey shaded region shows the approximate occurrence fractions of pulsar ESEs observed at lower frequencies predicted by our model as a function of CR mean free path. Because they are observed at lower frequencies, pulsars undergo strong scintillation at much higher rates than quasars observed at higher frequencies. Figure~\ref{fig:Wsrc_lambdaCR} shows that a CR mean free path of order $ 10$~pc is approximately consistent with the observed occurrence fraction of quasar ESEs. We note that in Figure~\ref{fig:Wsrc_lambdaCR} we have assumed that $\ell/s$ is sufficiently large such that $ \alpha_s > s/ D_\mr{lens}$ (eq.~\ref{eq:kappa}) and so equation~\eqref{eq:kappanumber} is valid for all of the chosen frequency bands.

\subsection{Constraints on sheet parameters} \label{sec:sheet_params}
In section~\ref{sec:rate_heuristic}, we showed that the observed mean free path of $\sim$GeV CRs can be reproduced using the observed occurrence fraction of ESEs and the sheet model considered in this work for density fluctuations similar to the warm ISM mean density ($n_e\sim0.03$cm$^{-3}$). 
In this section, we do not assume a value of $n_e$, and instead solve for all the sheet parameters simultaneously by incorporating more observational constraints. 
Here, we only keep the single basic assumption of this paper -- that CR scattering and ESEs are caused by the same density/current sheets. 

In Appendix~\ref{sec:gaussian_lens}, we provide a detailed calculation of the lensing behavior from an aligned sheet, assuming the electron column density variation introduced by the sheet can be approximated by a Gaussian function.  For $\kappa>2$, equation~\eqref{eq:wsrcnumber} above is a good approximation for both underdense and overdense sheets.

In addition to the occurrence fraction, there are two additional constraints from the ESE observations. The maximum bending angle and the demagnification region. The demagnification occurs when the source is directly behind the aligned sheet and the LOS flux is bent away from the observer (shown in Figure~\ref{fig:lensFreq}b). This region directly tracks the shape of the lens and hence is almost achromatic and can be approximated by
\begin{align}
    \beta_\mr{demag}=C_2 s/D_\mr{lens},
    \label{eq:beta_demag}
\end{align}
where $C_2$ is an order-unity constant that is again determined by the shape of the lens.

The right panel of Figure~\ref{fig:lensFreq} illustrates the modulation of radio flux with respect to source position and frequency, caused by an overdense aligned sheet.  The red dashed line is the approximation from equations~\eqref{eq:kappa},~\eqref{eq:kappanumber} and~\eqref{eq:beta_focal}. It nicely traces the outer boundary of the source position where the lens modulates the flux. 
While this lens-affected region is highly chromatic, the demagnification region introduced by the aligned sheet is almost achromatic for both over and underdense sheets. For a Gaussian lens, the demagnification can be approximated with: 
\begin{align} \label{demag_lens_ratio}
	\frac{\Omega_\mr{demag}}{\Omega_\mr{lens}}\sim \frac{\beta_\mr{demag} D_\mr{lens}}{s}\sim 3.
\end{align}
This is shown as the blue dashed line in Figure~\ref{fig:lensFreq}. 
The duration of demagnification in an ESE (labelled in Figure~\ref{fig:esetemperal}) can be estimated with: 
\begin{align}
    T_\mr{demag}=\frac{3 s}{D_\mr{lens}\omega_\perp}.
    \label{eq:Tdemag}
\end{align}
This constant demagnification region can be used to estimate the thickness of the sheet. 

Now we have four main observables: the CR mean free path $\lambda_\mr{CR}$ (equation~\ref{eq:CR_mfp_MW}), 
the ESE total duration (equation~\ref{eq:Tfocal}, Figure~\ref{fig:lensFreq} panel d), the demagnification time, as well as the ESE occurrence fraction (equation~\ref{eq:wsrcnumber}): 
\begin{align}
\begin{split}
     &\lambda_{\rm CR} 
= \frac{\ell}{f_v}\sim 1-10 \ {\rm pc}, \\ 
    &T_\mr{demag} \approx \frac{3s}{D_\mr{lens}\omega_\perp}\sim \mr{weeks} - \mr{months},\\
    &T_\mr{focal} \approx \frac{\kappa s}{D_\mr{lens}\omega_\perp}\sim \mr{weeks}- \mr{months}, \\    
    &W_\mr{src}\approx4\lambda^2 \frac{r_e}{2 \pi} D_{\rm lens} L \Delta n_e \frac{f_V}{\ell s}
    \sim 0.01.
    \label{eq:solve}
    \end{split}
\end{align}
$\omega_\perp$, the relative angular velocity of the source along the sheet normal, can be estimated from the full width at half maximum of the narrow peaks of the light curve, $T_\mr{peak}$, with: 
\begin{align}
     \omega_\perp = \frac{\Delta \beta_\mr{src}}{T_\mr{peak}}.
\end{align}
The duration of the peaks are expected to be the time a source with angular diameter $\Delta\beta_\mr{src}$ passes the infinitely narrow caustic region, and $\Delta \beta_\mr{src}$ is known from VLBI measurements. Alternatively $\omega_\perp$ can be estimated with $\omega_\perp = v \cos\phi/D_\mr{lens}$, where the relative velocity $v$ is usually $10-300$ km/s.

We can constrain the four main properties of the sheet via  
\begin{align} 
    &s\sim &&1\,\mr{AU} \frac{T_\mr{demag}}{\mr{month}}\frac{D_\mr{lens}}{\mr{kpc}}\frac{\omega_\beta}{0.1\mr{mas/day}},\nonumber\\
    &\frac{s}{l}\sim &&3\times10^{-5}\,\frac{T_\mr{demag}}{T_\mr{focal}}\left(\frac{L}{\mr{kpc}}\right)^{-1} \frac{W_\mr{src}}{10^{-2}}\frac{\lambda_\mr{CR}}{10\mr{pc}},\\
    &f_V\sim &&0.01  \frac{D_\mr{lens}L}{\mr{kpc}^2}
    \frac{T_\mr{focal}}{\mr{month}}
    \frac{\omega_\beta}{0.1\mr{mas/day}} 
    \left(\frac{W_\mr{src}}{0.01}\right)^{-1} \left(\frac{\lambda_\mr{CR}}{10\mr{pc}}\right)^{-2},\nonumber\\
    &\Delta n_e \sim&& 0.03\,\mr{cm^{-3}}\nonumber\\
   & &&\, \frac{T_\mr{demag}}{\mr{month}}\frac{\omega_\beta}{0.1\mr{mas/day}}
    \frac{W_\mr{src}}{10^{-2}}\frac{\lambda_\mr{CR}}{10\mr{pc}}\left(\frac{L}{\mr{kpc}}\right)^{-1} \left(\frac{\nu}{3\mr{GHz}}\right)^2\nonumber.
\end{align}
The resulting sheet parameters are consistent with those proposed for ESE events alone when assuming pressure equilibrium (see summary in \citealt{2018Vedantham}). The electron density is comparable to order of magnitude to the average 
$n_e$ in the ISM, thereby avoiding the issue of overpressure. This large axial ratio has also been proposed to explain the scintillation of PSR B0834+06 \citep{Simard:2018}. Whether such large aspect ratios are indeed produced on small scales in interstellar turbulence remains to be tested. The estimated sheet parameters also give two additional observables, 
\begin{align}
\begin{split}
        \Delta \mr{DM} &\sim \Delta n_e l \sim  10^{16}\,\mr{cm}^{-2} \frac{n_e}{0.1\mathrm{cm}^{-3}}\frac{\ell}{10^4\mathrm{AU}}, \sim0.005\,\mr{pc}\,\mr{cm}^{-3} \\
        \Delta \mr{RM} &\sim \Delta n_e l B \sim  0.05\,\mathrm{rad}\,\mr{m}^{-2}\frac{n_e}{0.1\mathrm{cm}^{-3}}\frac{\ell}{10^4\mathrm{AU}}\frac{B}{10\mu G}.
        \end{split}
\end{align}
The interstellar magnetic field is of order $\sim 1-10 \ \mu$G, however, we note that the magnetic field inside the sheet drops in magnitude as it reverses direction, so the $\Delta$RM may be an overestimate. It is worth noting that the constrained sheet number density, $n_s\sim f_V/(\ell^2 s)\sim 10^5 \ \mr{pc}^{-3}$, is much greater than the stellar density in the galactic disk, $ n_* \approx 0.1 \ \mr{pc}^{-3}$. 

 \section{Observational predictions}\label{sec:obs_predictions}
Observations of CR secondary-to-primary ratios, such as the boron-to-carbon ratio, imply very long lifetimes of CRs in the ISM, orders of magnitude longer than the Milky Way's light-crossing time. Obtaining direct information on the scattering structures from observations of CRs is thus challenging because they undergo many scattering events throughout the Galaxy before reaching Earth. In contrast, radio waves retain information about their propagation paths, making them valuable for probing underlying structures. Since GeV CRs may be scattered strongly by magnetic current sheets with AU-scale thicknesses, confirming the existence and characteristics of these sheets through radio wave observations is essential, and offers a novel way to constrain CR propagation. In this section, we propose observational predictions and tests for the current-sheet model. With upcoming wide-field radio survey instruments, such as DSA 2000 and CHORD, there are ample opportunities to collect a large number of ESE observations, enabling detailed statistical analyses to confirm or refute this model.

\begin{figure*}
\includegraphics[width=\linewidth]{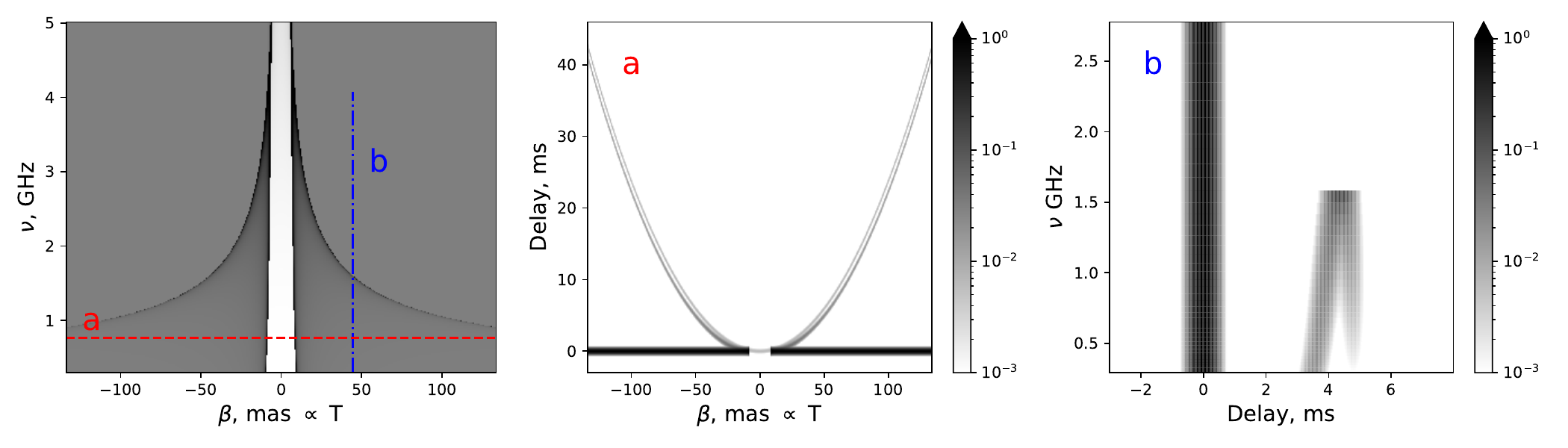}
\vspace{-0.1cm}
\caption{ESE behaviors for short-duration radio transients. Large bending angles of the lensed images induce significant time delays, enabling separation by arrival time. This allows the study of lensing structures by tracking the temporal and spectral evolution of individual images, rather than relying on the total flux as done for quasars. (Left): The total flux variation summed from all images, similar to Figure~\ref{fig:lensFreq}. (Panel a): Variation of brightness (greyscale) and time delay between different images over time as the source moves, sampled at the frequency indicated by the red horizontal dashed line in the left panel. (Panel b): Brightness of the images (greyscale) as a function of frequency and delay, measured at the position of the blue vertical dashed line in the left panel. This figure is generated assuming the Gaussian overdense lens described in Appendix~\ref{sec:gaussian_lens} with $\kappa=10(\nu/3\mr{GHz})^{-2}$. }
\label{fig:ESEtransient}
\end{figure*}

\subsection{ESEs for pulsars and fast radio bursts (FRBs) at lower frequency}
\label{subsec:sheetparameter}
The detection of ESEs with quasars can be distorted by intrinsic light curve modulation. It would be valuable to confirm the rate and lensing behaviors with transients like FRBs and pulsars, where the ESEs can show up as echos of the main images.

The occurrence fraction for observing a lensed image is proportional to the largest bending angle of the sheet (equation~\ref{eq:wsrcnumber}), and hence: 
\begin{align}
    W_\mr{src}(\nu)\sim10^{-2} \left(\frac{\nu}{3\mr{GHz}}\right)^{-2},
\end{align}
where $W_\mr{src} (3 \ \mr{GHz})\sim 10^{-2}$ is measured in \citealt{1987Fiedler,1994Fiedler}. 
The magnification zone on the source plane gets very extended at low frequencies, but the level of magnification decreases. The demagnification zone is achromatic, therefore the reduction of flux remains significant and the demagnification occurrence fraction $W_\mr{demag}$ is:
\begin{align}
    W_\mr{demag}(\nu)\sim10^{-3}-10^{-2}.
\end{align}

The bending angle of the lensed image can be as large as
\begin{align}
    \alpha\sim 17\mr{mas}\left(\frac{\nu}{\mr{GHz}}\right)^{-2}
    \left(\frac{\Delta \mr{DM}}{0.005\mr{pc}\,\mr{cm}^{-3}}\right)
	\left(\frac{s}{\mr{AU}} \right)^{-1}.
\end{align}
At 1 GHz, the images can be bent by more than 10 mas, which is resolvable with Very Long Baseline Interferometers (VLBI). By performing several VLBI observations across different weeks, it should be possible to track the evolution of the image separations, which would in turn help constrain the sheet structures. 

The arrival time difference of the lensed and unlensed images will also be significant: 
\begin{align}
    \tau&\sim D_\mr{lens}\alpha^2/c\\
    &\sim 0.7\mr{ms}\left(\frac{\nu}{\mr{GHz}}\right)^{-4}
    \left(\frac{\Delta \mr{DM}}{0.005\mr{pc}\,\mr{cm}^{-3}}\right)^2
	\left(\frac{s}{\mr{AU}} \right)^{-2}\frac{D_\mr{lens}}{\mr{kpc}}.\nonumber
\end{align}
The delay increases rapidly with decreasing frequency. For observations of transients above 1GHz, it is easier to look for ESEs from the flux modulation. This can be done with sources of known flux, such as quasars or pulsars. 
For observations below 1 GHz, quasars are no longer point-sources and hence do not show ESE effects. One can then look for ESEs by looking for the demagnification in pulsar flux, as well as searching for faint echos in pulsars and fast radio bursts (FRBs) if the delay time is comparable to or larger than the pulse duration of the pulsars ($10^{-2} - 10^3$ ms depending on the pulsar) or the duration of FRBs (usually ms). As shown in Figure~\ref{fig:ESEtransient}, the echos can have different frequency structure and are only observable for bright sources, as the lensed image can be $\sim10-1000$ times fainter than the unlensed image. The echo should be stable over the timescale of a day and gradually evolve on week to month timescales. Therefore, it is recommended to combine all the bursts and pulses detected within a few days to find the faint echos. 

For the Canadian Hydrogen Intensity Mapping Experiment (CHIME) and, in the future, the Bustling Universe Radio Survey Telescope for Taiwan (BURSTT), which operates at 600 MHz, approximately a quarter of the sources will exhibit lensed images produced by the sheets. The Deep Synoptic Array 2000-antenna (DSA-2000), operating between 0.7 and 2 GHz, will be capable of monitoring a large number of quasars and pulsars simultaneously, providing extensive statistics on the occurrence fraction of such phenomena. The Canadian Hydrogen Observatory and Radio Transient Detector (CHORD), with a frequency range of 300 to 1500 MHz, will detect all sources with echoes at the lower end of the band, enabling simultaneous studies of images across frequency and time, and generating maps of Extreme Scattering Events (ESEs), similar to those shown in Figure~\ref{fig:lensFreq}. 

\begin{figure}
    \centering
    \includegraphics[width=0.7\columnwidth]{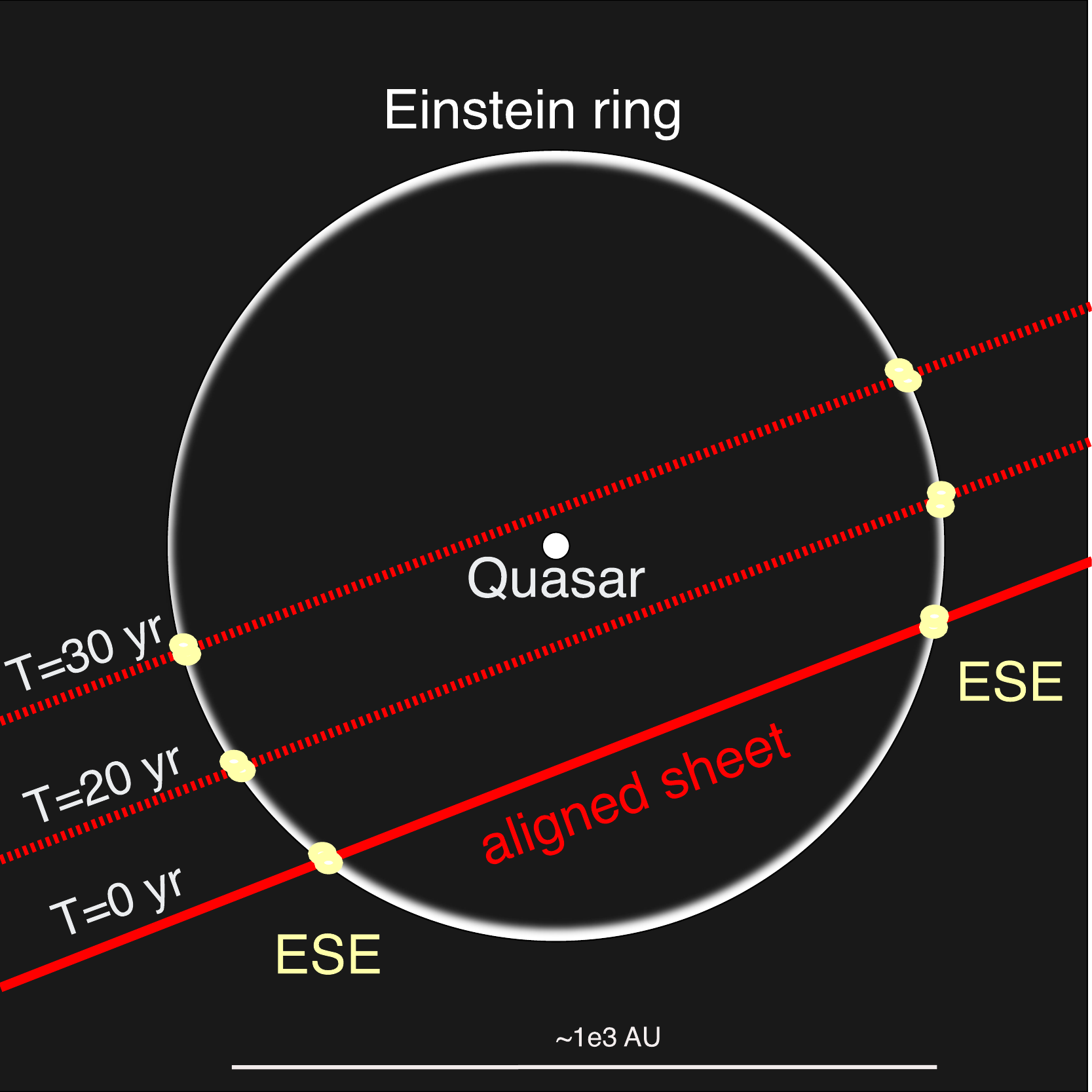}
    \caption{The Einstein ring of a gravitationally lensed quasar can serve as a probe of interstellar density sheets. In the presence of a sheet characterized by a large aspect ratio, two widely separated regions on the ring can simultaneously undergo an ESE. Such ESE pairs can be identified through synchronized variations in their temporal evolution.}
    \label{fig:einsteinRing}
\end{figure}

\subsection{ESEs in Einstein rings of quasars}
\begin{figure}
\begin{minipage}{0.3\linewidth}
\includegraphics[width=\columnwidth]{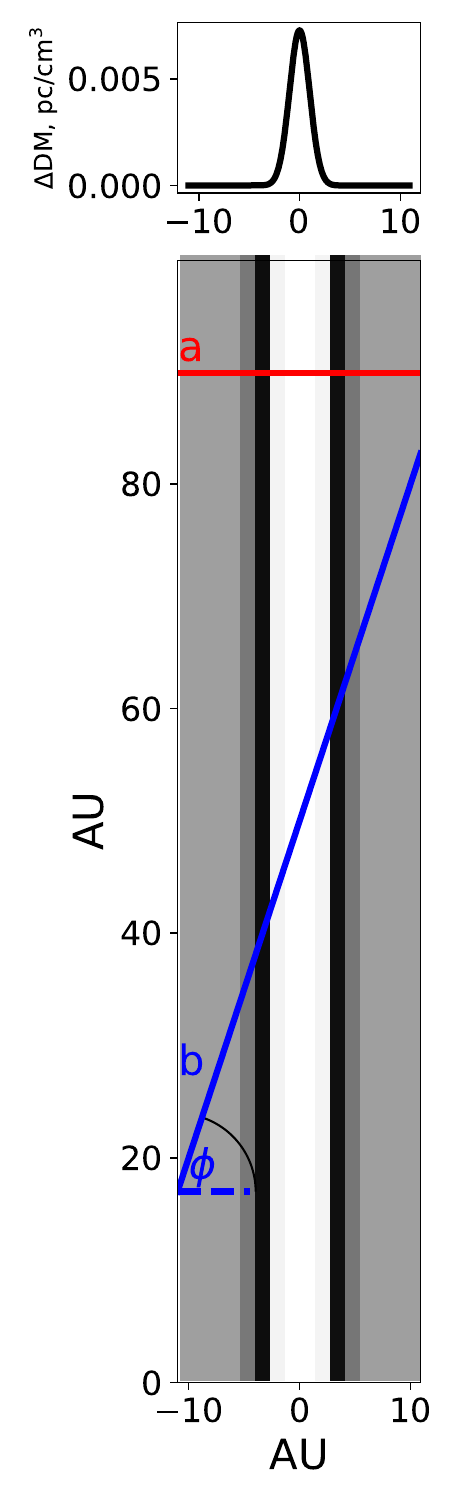}
        \end{minipage}
\begin{minipage}{0.6\linewidth}
\includegraphics[width=\columnwidth]{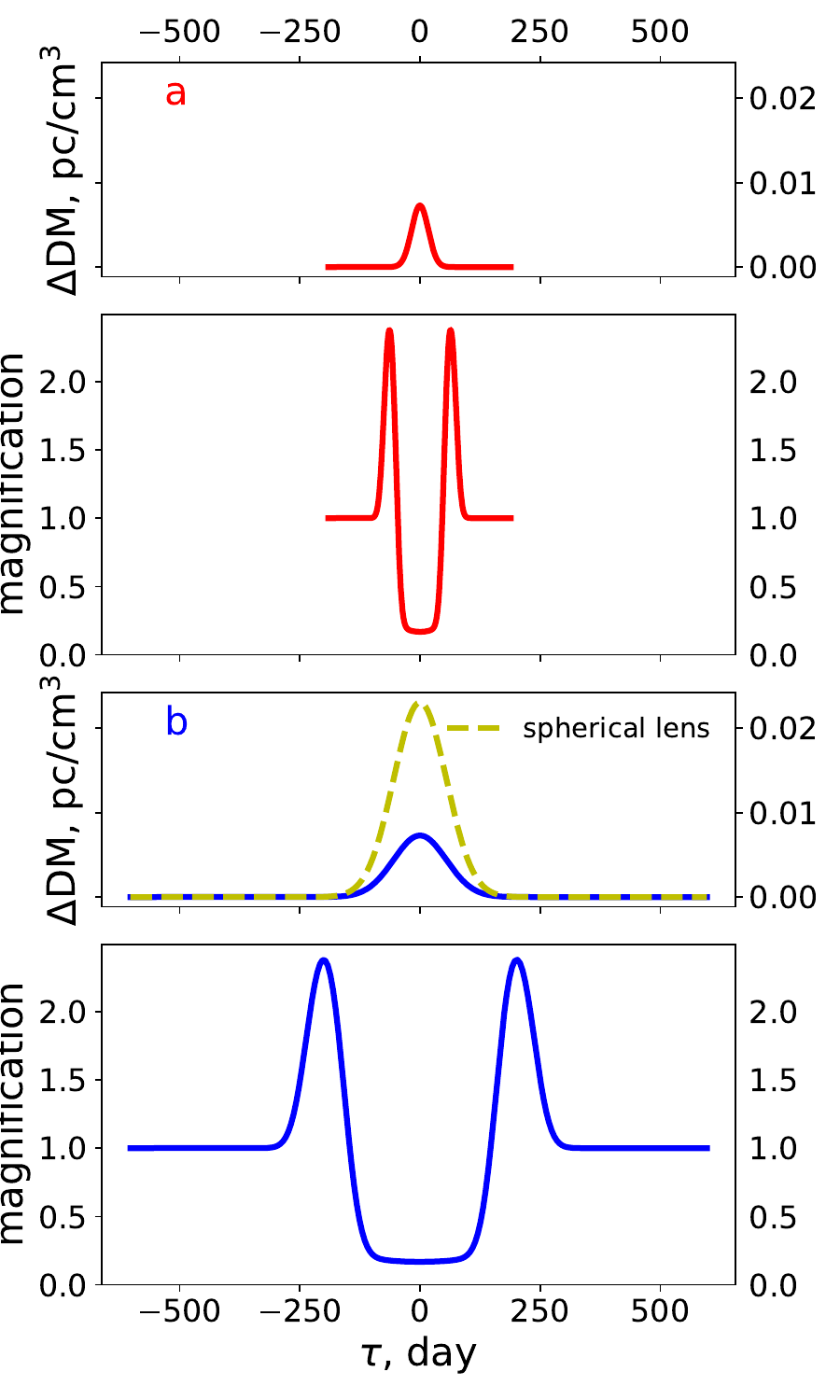}
        \end{minipage}

\caption{The duration of an ESE depends on the angle $\phi$ at which the source crosses the sheet. Left: the magnification of the source at different locations, projected on the lens plane (greyscale shading). The aligned sheet is located at $x=0$, with a DM change shown in the top panel. Two background sources, a and b, pass the sheet at different angles (red and blue lines). Right: the observed DM change (top) and magnification (bottom) of the sources a and b. The source that passes the sheet with larger $\phi$ shows a longer duration ESE. A much larger DM fluctuation would be necessary to produce long-duration ESEs using a spherical lens rather than a sheet, as shown by the yellow dashed line in panel b. \label{fig:passAngle}}
\end{figure}

An aligned sheet with parameters estimated in section~\ref{sec:sheet_params} has a highly elongated focal region in the source plane where flux modulation is anticipated (Figure~\ref{fig:passAngle} left panel): 
\begin{align}
	\Omega_\mr{focal}\sim& 0.01^{''}\frac{\kappa s}{10\mr{AU}}
	\left(\frac{\nu}{3\mr{GHz}}\right)^{-2}\left( \frac{D_\mr{lens}}{\mr{kpc}}\right)^{-1}\\
    &\times 10^{''}\frac{l}{10^4\mr{AU}}
	\left( \frac{D_\mr{lens}}{\mr{kpc}}\right)^{-1} \nonumber\\
	\sim& 0.1\, \mr{square-arcsecond} .\nonumber
\end{align} 
This geometry of the sheet can be probed by observing multiple sources from the focal region simultaneously. The Einstein ring of a strongly lensed quasar is a particularly nice probe of the highly elongated focal region introduced by the sheet. 

When a background quasar is located at the focal point of a foreground galaxy, an Einstein ring is produced. The angular radius of the ring can be estimated with, 
\begin{align}
	\theta_E=\left(\frac{4G M_f}{D^g_\mr{rel}\,c^2}\right)^{-1/2}
	=3^{''} \left(\frac{M_f}{10^{12}M_\mr{sun}}\right)^{-1/2}\left(\frac{D^g_\mr{rel}}{\mr{Gpc}}\right)^{1/2}.
\end{align}
Where $D^g_\mr{rel}$ is the relative distance defined as $[D_\mr{rel}^{g}]^{-1}=[D_\mr{lens}^{g}]^{-1}-D_\mr{src}^{-1}$, where $D^g_\mr{lens}$ and $D_\mr{src}$ are the distance from the observer to the lensing galaxy and to the source, respectively. 
We stress that while the radius of the Einstein ring at the location of the lensing galaxy is of order 10 kpc, $\theta_E D^g_\mr{lens}\sim 10\, \mr{kpc}\, (\theta_E/3^{''}) (D^g_\mr{lens}/\mr{Gpc})$, the projected radius of the ring at the location of the ESE-inducing-sheets in the Milky Way is around $10^3$ AU: $\theta_E D_\mr{lens}\sim 3\times 10^{3}\, \mr{AU}\, (\theta_E/3^{''}) (D_\mr{lens}/\mr{kpc})$. 
Therefore it is comparable to the long axis of the sheet. The thickness of the ring corresponds to the angular size of the quasar ($\sim$mas), which is similar to the sheet thickness, $s$, and hence thin enough for ESEs to happen. 

As shown in Figure~\ref{fig:einsteinRing}, when an elongated sheet passes in front of the Einstein ring, we can expect two points experiencing an ESE simultaneously. In 10 years, the relative position of the sheet and the ring will move by $\Delta \theta = 0.4^{''}\, (T/10\,\mr{yr}) (\omega_\beta/0.1\mr{mas}/\mr{day})$. Therefore two new points on the ring will experience an ESE. 
Observations of two ESE points with correlated movement against time would be consistent with ESEs produced by a thin sheet. The relative movement of the two points would in turn constrain the orientation and radius of curvature of the sheets. Einstein rings therefore offer a unique opportunity to directly measure the large axial ratios (down to $10^{-4}- 10^{-3}$) of the sheets from the separation of the two ESE points and the size of the ESE points. 

The mean free path of a point source moving on the sky without encountering an ESE from the foreground sheets can be estimated using the rate of ESEs (e.g. \citealt{1987Fiedler}): 
\begin{equation}
	\lambda_\mr{src}\sim \frac{\kappa s} {W_\mr{lens}}\sim  10^3\,\mr{AU}\frac{\kappa s}{10 \ \mr{AU}}\left(\frac{W_\mr{lens}}{10^{-2}}\right)^{-1},
\end{equation}
and the corresponding angular mean free path is 
\begin{align}
\lambda^\theta_\mr{src}\sim \frac{\lambda_\mr{src}}{D_\mr{lens}}=1^{''}\frac{\kappa s}{10 \ \mr{AU}}\left(\frac{W_\mr{lens}}{10^{-2}}\frac{D_\mr{lens}}{\mr{kpc}}\right)^{-1}.
\end{align}
Therefore, for an Einstein ring that spans several arcsecond,  there is high probability that some of the points will encounter ESEs. In fact, it is likely to encounter several aligned sheets in the foreground, and so multiple measurements of the points variation against time would be required to solve for correlated ESE behaviors. 

Very thin Einstein rings from quasars have been observed for at least two cases: Quasar B1938+666 and J0751+2716 \citep{2018Spingola}. The rings are incomplete for both cases, and this reduces the chance of seeing correlated points by a factor of $f=(\Delta \theta_\mr{ring}/2\pi)^2$. The analysis we propose here is still possible with enough measurements against time. 
In the near future, DSA-2000 is expected to detect $\sim 10^4$ lensed AGN, which will provide a good sample to probe intermittent sheets in the ISM. 

\subsection{DM structure function}
When emission from a background pulsar passes through an aligned current sheet, it will encounter a small change in dispersion measure, $\Delta$DM$=\int \Delta n_e dl\sim 10^{-3}-10^{-2}$ pc/cm$^3$. The crossing takes $T\sim 0.5\,\mr{month}(s/\mr{AU})(v/100\mr{km}/s)^{-1}$, and the chance of alignment is $W_\mr{lens}\sim10^{-3}$ (equation~\ref{eq:wl}), independent of the observation frequency. The $\Delta$DM from the aligned sheets can show up in the DM structure function. Thus, one can use measurements of the DM structure function as an additional probe of interstellar current sheets that is independent of the lensing modeling. 

The DM structure function can be measured by observing the pulsar DM variation against time:
\begin{align}
    D_\mr{DM}(\Delta x)=D_\mr{DM}(\tau)=\langle [\mr{DM}(t)-\mr{DM}(t+\tau)]^2 \rangle,
    \label{eq:DMstruct}
\end{align}
where $\tau=\Delta x/v$ and $v$ is the relative velocity. For aligned sheets, the  decorrelation time $\tau_\mr{decor}\sim s/v\sim 0.5 \,\mr{month}\,(s/\mr{AU})(v/100\mr{km}/s)^{-1}$, therefore it will introduce additional power in the structure function on timescales of days to months. The AU-thick aligned sheet estimated from ESEs would introduce:
\begin{align}
    &\Delta D_\mr{DM}^\mr{sheets}(\tau\sim\mr{days - months})\sim W_l \Delta \mr{DM}^2 \\
    &\sim 10^{-7}\, \mr{pc}/\mr{cm}^3\, \frac{W_\mr{lens}}{0.002}\left(\frac{\Delta\mr{DM}}{0.005\mr{pc}/\mr{cm}^3}\right)^2. \nonumber
\end{align}
For sheets with different thicknesses, the relative volume filling factor could be probed using the variation of structure function against $\tau$ and equation~\ref{eq:wl}:
\begin{align}
    \Delta D_\mr{DM}^\mr{sheets}(\tau)&\sim \int_\mr{s_\mr{min}}^{v\tau} W_\mr{lens} \Delta \mr{DM}^2 ds \,\\
    &\sim L\Delta n_e^2 \int_\mr{s_\mr{min}}^{v\tau} f_V(s,l) s ds,\nonumber
\end{align}
where $s_\mr{min}$ is the smallest sheet thickness. For $f_v\propto s^{-2}$ and $\ell$ independent of $s$, the structure function has equal contributions from all logarithmic $s$ bins. 

For estimating the structure function, one important thing to notice is the amount of time needed to achieve statistics good enough to perform the averaging in equation~\eqref{eq:DMstruct}. For a given source, it takes on average $T=s/W_\mr{lens}/v \sim 50\,\mr{yr}$ to pass by an aligned sheet. Therefore, if the structure function is measured over times much shorter than this timescale, we expect to see no contribution from sheets in most cases, and much larger disturbances for a small fraction of the cases. A very stable increase in the DM structure function with data spanning only a few years would indicate a different origin. 

For unaligned sheets, the decorrelation time $\tau_\mr{decor}\sim l/v_\mr{rel}\sim 500 \,\mr{yr}\,(l/10^4\mr{AU})(v/100\mr{km}/s)^{-1} $, which is well beyond the current observational time limits and hence would not affect the DM structure function measured over a hundred years. 

\subsection{Long-duration ESEs}
The duration of an ESE depends on the angle $\phi$ between the velocity of the source and the direction of the sheet normal: $T\propto s/\cos{\phi}$ (Figure~\ref{fig:passAngle}). 
For a source moving in the plane of the sky, the probability of encountering an aligned sheet anywhere along the LOS with projected angle $\phi$ is $P(\phi)\propto l\cos{\phi}\,T \sim ls$, which is independent of $\phi$. Therefore, for a population of sources currently experiencing ESEs, 5\% of them will be passing the sheet with $\phi$ between $85^\circ$ and $90^\circ$, and experiencing an ESE that's 10 times longer than for $\phi\sim 0 ^\circ$. The ESE behaviors for $\phi=0^\circ$ (red lines) and $70^\circ$ (blue lines) are shown in Figure~\ref{fig:passAngle}. For the trajectory with larger $\phi$, the duration of the ESE is three times longer. The peak of the observed DM variation would stay the same but the change will be observed on a three times longer timescale. When the Earth motion dominates the relative motion, reoccurrence of ESE events may be seen with the separation of roughly a year. 

\subsection{Nearly parallel sheets and implications for ESEs}
The sheets derived in Section~\ref{sec:sheet_params} have a volume filling fraction of $\sim 1\%$ and an aspect ratio of $\sim 10^{4}$. The average spacing between adjacent sheets is therefore smaller than the lengths of their long axes. Neighboring sheets thus need to be almost parallel to avoid field lines crossing each other. 
Moreover, since the constrained volume filling fraction is an ISM-averaged value, it is likely that there exist regions with a higher sheet filling fraction and voids where there are very few such coherent structures. In regions with higher filling fraction, the sheets would therefore be layered more tightly and be even closer to parallel.
This picture is supported by Figure~\ref{fig:dNedx}, particularly the right panel, where neighboring structures that are very close to each other are nearly parallel.
Observationally, sources passing nearby parallel sheets are expected to experience frequent ESEs, like in the case of PSR 1937+21. For such sources, VLBI images from different ESEs are predicted to show a preferred offset direction in our model.

\subsection{Environmental dependence of ESEs}

ESEs require thin sheets with large aspect ratios and significant density variations. The existence of such structures can be facilitated by the presence of an enveloping magnetic field that reverses direction and keeps the system in pressure balance (\citealt{2006GoldreichSridhar}; Figure~\ref{fig:scattering_schematic}). At the same time, the field reversal acts as a source of CR scattering. This configuration, however, requires that there exist magnetic-field reversals on very small, $\sim$AU, scales. Recent attempts to  characterize the intermittency of MHD turbulence as a function of the properties at the driving scale (typically $\sim 10-100$~pc in the ISM) demonstrated that such field reversals likely exist on all scales (with scale-dependent volume-filling fractions) if the turbulence is driven super-Alfv\'enically (\citealt{Yuen:2020}; \citealt{Fielding:2023}; \citealt{Kempski:2023}), i.e. if the turbulence at the outer scale is strong enough to significantly perturb the large-scale ordered field $B_0$ and produce $\delta B/B_0 \gtrsim 1$. Thus, if ESEs and CR scattering are indeed due to intermittent sheet-like structures produced in interstellar turbulence, one might expect higher rates in regions of high turbulent activity. Because of the relatively small CR mean free path, observations of localized enhanced CR scattering are challenging. The small occurrence fraction of ESEs make them a better candidate for testing correlations with turbulent amplitudes. The magnitude of turbulent magnetic fields in the Milky Way can be estimated from RM fluctuations, and there are regions in the galactic plane that show large variations in observed pulsar RMs (\citealt{Haverkorn:2015}). Comparing directions/locations of ESEs with available RM maps (e.g., \citealt{Hutschenreuter:2022}) may give new insights into the types of turbulence and intermittency that give rise to ESEs and provide a test for the turbulent sheet model, which predicts that ESEs should preferentially occur in regions with large fluctuating magnetic field components.

\section{Discussion} \label{sec:discussion}

\subsection{Galactic regions probed by CRs versus ESEs}
Our work implicitly assumes that the statistics of magnetic-field reversals are comparable in the warm ISM and inner CGM. However, the density fluctuations resulting from field reversals are expected to differ significantly between these two environments. While CRs are scattered in both regions, the large column density gradients required for ESEs likely only occur in the warm ISM. Therefore, the Galactic regions probed by observations of CRs and ESEs are generally not the same. In particular, the $\sim$GeV CR mean free path of order $1-10$~pc that we adopted in this paper is consistent with observed CR spectra assuming that CRs spend most of their galactic residence time in the diffuse gas filling the inner CGM. By contrast, observed ESEs are most prevalent in the inner 1~kpc of galactic latitudes, where the warm ionized medium occupies an order-unity volume-filling fraction. Moreover, there is tentative evidence, with $2.5\sigma$ significance, that ESEs tend to occur near the edges of galactic continuum loops \citep{1994Fiedler}, which are typically linked to supernova remnants, Galactic outflows, or OB star complexes. Further away from the disk, tenuous hot gas becomes the volume-filling phase, with densities too low to produce significant radio-wave scattering even if similar magnetic structures exist that will scatter CRs. The CGM also hosts cooler and denser clouds, which can scatter radio waves (\citealt{VedanthamPhinney2019}), however, those will not dominate ESEs.
In the future, it should be possible to measure the scale-height dependence of the sheets by determining the lens distance of a large number of ESEs with VLBI. It would also be helpful to study the scattering of quasars and FRBs with different impact parameters to the disk of a foreground galaxy.

\begin{figure*}
  \centering   \includegraphics[width=\textwidth]{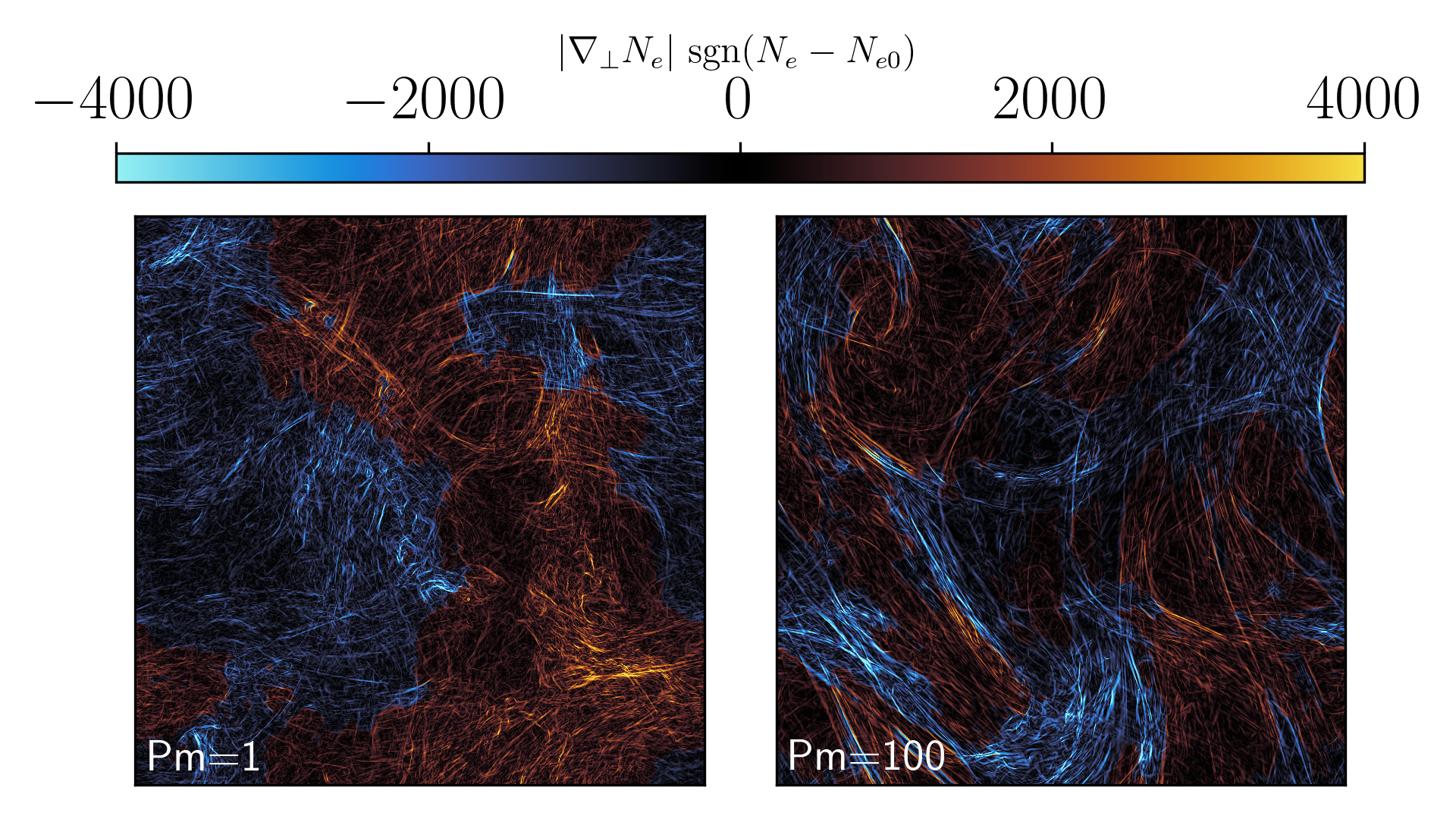}
  \caption{Magnitude of the in-plane gradient of projected column density (arbitrary units) in two high-resolution ($2240^3$) MHD dynamo simulations from \citealt{Galishnikova:2022}. Blue denotes regions where the local column density $N_e$ is less than the mean column density $N_{e0}$, while red and yellow regions correspond to locally enhanced column density. The simulation shown in the left panel was run with Pm=1 (smaller viscosity) while the simulation shown in the right panel was run with Pm=100 (larger viscosity). Both simulations use the same resistivity. Notably, in both panels we see that regions of large column density gradients are highly intermittent and that there are numerous sheets that are remarkably straight. The Pm=1 case shows appreciably larger contributions from sheets with shorter long-axis lengths relative to the Pm=100 case, demonstrating that how turbulence is dissipated has significant implications for the geometry of density sheets and thus the occurrence of ESEs.}
  \label{fig:dNedx}
\end{figure*}

\subsection{Sheet geometry and the role of dissipation \label{sec:dissipation}}
Throughout our calculation we have assumed that the density/current sheets are very straight. While this picture is arguably theoretically more plausible than the alternative of quasi-spherical, over-pressurized, high-density lenses being responsible for ESEs, we revisit our assumption here. In particular, we consider \textit{how straight} the sheets need to be for our preceding calculations to be valid. One can show that for a sheet with dimensions ($\ell$,$\ell$,$s$) but with finite global radius of curvature $r_c$, the maximum path length a light ray can propagate along the sheet is of order $(r_c s)^{1/2}$ and the chance of such alignment is $\sim \ell / r_c$ (\citealt{2006GoldreichSridhar}). This means that, for a sheet with $\ell \gg s$ to be considered effectively straight, one requires an extremely large radius of curvature, $r_c \gtrsim \ell^2 / s$. Such large radii of curvature are also necessary not to overproduce the rates of ESEs, given the high alignment chance $\sim\ell / r_c$ and the observed CR mean free path, and adopting the unified scattering picture depicted in Figure~\ref{fig:scattering_schematic}.\footnote{These estimates are further modified by additional geometric considerations such as sheet corrugations existing on all scales or unequal long axis ratios ($\ell_1 \neq \ell_2$), which would in turn act to reduce the occurrence fraction of ESEs.} We discuss alternative models of CR and radio-wave scattering by sheets in Appendix~\ref{sec:alternative_sheet}, which result in different geometrical constraints on the sheet parameters.

Given the numerous geometric degrees of freedom, generalizing our calculation to more complicated sheet geometries is beyond the scope of this work. While the assumption of extremely straight sheets may seem highly artificial at first glance, this geometry is favoured by observations of ESEs and pulsar scintillation (\citealt{2010Brisken}; \citealt{2023Koryukova}). Moreover, for a fixed radius of curvature $r_c$ and corresponding maximum radio-wave path length within the sheet of order $\sim (r_c s)^{1/2}$, the observationally inferred column density gradients of order $dN_e / d x \gtrsim1000 \ {\rm cm}^{-3}$ require $r_c \gtrsim 10^8 s$ for characteristic density variations of order $0.1 \ {\rm cm}^{-3}$. Sheets with order-unity global curvature, i.e.~$\ell \sim r_c$, would therefore need to have extreme aspect ratios of order $\ell/s \gtrsim 10^8$ to explain the observed large column density gradients. Both candidates for producing large scattering, smaller-aspect-ratio almost-straight sheets and extreme-aspect-ratio curved sheets thus need to satisfy stringent geometric constraints.\footnote{However, it is worth noting that current sheets formed in MHD turbulence simulations often show remarkably large global radii of curvature (Figure~\ref{fig:dNedx}).} Interestingly, even if we consider the straight sheets with $\ell / s \sim 10^4$ to be a subset of a population of sheets with varying radii of curvature, as long as the straight sheets have a large enough volume filling fraction they may be enough to explain the observed CR scattering and quasar ESE rates (Figure~\ref{fig:Wsrc_lambdaCR}). On the other hand, if the vast majority of current sheets do not satisfy $r_c \gg \ell$, then the occurrence fraction and bending angles of ESEs require sheets with very large aspect ratios but much smaller volume filling fractions, implying CR mean free paths due to these structures that are much larger than observed.   

Interstellar turbulence is a plausible mechanism for the formation of the sheets considered in this work. While it is not clear what physical mechanisms would be responsible for the large radii of curvature of current/density sheets producing ESEs, it is likely that the microphysics of how turbulence is dissipated plays an important role. Intriguingly, in the warm ionized medium the thermal ion mean free path is comparable to the thickness of sheets producing ESEs,
\begin{equation}
    \lambda_{\rm ii} \sim 1 \ {\rm AU} \ \Big(\frac{T}{10^4 \ {\rm K}}\Big)^{2} \Big(\frac{n_i}{0.03 \ {\rm cm}^{-3}}\Big)^{-1},
\end{equation}
where $n_i$ is the ion number density. For a thermal speed $v_{\rm th}$, this mean free path with associated viscosity $\nu_\mr{B} \sim v_{\rm th}\lambda_{\rm ii} $ corresponds to a Kolmogorov microscale
\begin{equation}
    \ell_{\nu} \sim 100 \ {\rm AU} \  \Big( \frac{L_{\rm T}}{100 \ {\rm pc}} \Big)^{1/4} \Big(\frac{\lambda_{\rm ii}}{1 \ {\rm AU}} \Big)^{3/4} \ \Big(\frac{\rm Ma}{0.5}\Big)^{-3/4},
\end{equation}
where $L_{\rm T}$ is the turbulence driving scale and ${\rm Ma}$ is the sonic Mach number. While this dissipation scale is computed for an isotropic hydro-like cascade and isotropic viscosity, rather than anisotropic MHD turbulence characterized by anisotropic (Braginskii; \citealt{Braginskii:1965}) viscosity, this simplified calculation nevertheless demonstrates that ESEs and CR transport probe scales on which dissipation of turbulence becomes important (although we note that the anisotropic viscous stress acts primarily in the direction of the local magnetic field, which is along the long axis of the sheet and not the short axis). This will plausibly change the morphology and geometry of current sheets produced by the cascade. We illustrate this dependence on dissipation in Figure~\ref{fig:dNedx}, which shows the magnitude of the 2D (in-plane) gradient of projected column density      computed from two snapshots of the MHD dynamo simulations run by \citealt{Galishnikova:2022}. The magnitude of the column density gradient is of direct relevance to ESEs as it determines the bending angle due to density fluctuations (eq.~\ref{eq:alpha}). Both simulations shown in Figure~\ref{fig:dNedx} were run at resolution $2240^3$ with resolved dissipation and the same value of resistivity. The simulation shown in the left panel was run with  magnetic Prandtl number ${\rm Pm} = \nu_\mr{B}/\eta =1$ (where $\nu_\mr{B}$ is the viscosity and $\eta$ is the resistivity) while the simulation shown in the right panel was run with ${\rm Pm}=100$ (larger viscosity). Notably, in both panels we see that regions of large column density gradients are highly intermittent and that there are numerous sheets that are remarkably straight. The ${\rm Pm}=1$ case shows appreciably larger contributions from sheets with shorter long-axis lengths relative to the ${\rm Pm}=100$ case, demonstrating that how turbulence is dissipated has significant implications for the geometry of density sheets and thus the occurrence of ESEs. Future simulations with more realistic anisotropic viscosity will be essential for understanding the geometry of current/density sheets formed in realistic interstellar turbulence. In particular, recent work has shown that anisotropic viscous forces cause turbulent plasmas to re-organize their flow to minimize changes in magnetic-field strength (\citealt{Squire:2019}; \citealt{Squire:2023}; \citealt{Majeski:2024}), a fundamental difference from ideal MHD turbulence. This ``magneto-immutability" of weakly collisional plasmas may play a key role in how current sheets are formed in interstellar turbulence. Finally, we note that the fact that GeV CRs are energetically important in the ISM may also affect the geometry of current sheets on scales comparable to their gyroradii in non-negligible ways. 

\subsection{Disruption by tearing and sheet lifetime \label{sec:tearing}}
Equation~\ref{eq:wsrcnumber} suggests that the occurrence fraction of ESEs is expected to increase with decreasing sheet thickness $s$. However, observations of ESEs are predominantly due to $\sim$AU-scale lenses, suggesting that scattering by smaller-scale structures is suppressed. For quasar ESEs, the lack of scattering by sheets much thinner than an AU can be explained by the large angular size of the source (\citealt{Stanimirovic:2018}). However, for pulsars thinner sheets should be the dominant scatterers, unless they have significantly smaller volume filling fractions. One possibility is that the formation of elongated sub-AU sheets is suppressed by their disruption due to tearing. If the sheets are a product of interstellar turbulence, their minimum thickness can be estimated by balancing the tearing timescale with the turbulent decorrelation timescale. \cite{Galishnikova:2022} estimate this characteristic perpendicular reversal scale in the turbulent MHD dynamo to be
\begin{equation} \label{eq:s_dynamo}
    s \sim L_{\rm T}\, {\rm Rm}^{-3/10} {\rm Pm}^{-1/5},
\end{equation}
where ${\rm Rm} = UL_{\rm T}/\eta$ is the magnetic Reynolds number and $U$ is the characteristic velocity at the driving scale of the turbulence. (Here, we have assumed that the sheets are more sinusoidal in their profiles than tanh-like -- see eq.~12 of \citealt{Galishnikova:2022}.) Using Spitzer values for the kinematic viscosity and resistivity of a fully ionized plasma, we find,
\begin{align} \label{eq:s_ism}
    s & \sim   0.2 \ {\rm AU} \ \Big( \frac{L_{\rm T}}{100~{\rm pc}} \Big)^{7/10} \Big( \frac{U}{10~{\rm km/s}} \Big)^{-3/10} \\ & \times \Big( \frac{n_i}{0.1~{\rm cm^{-3}}} \Big)^{1/5} \Big( \frac{T}{10^4~{\rm K}} \Big)^{-5/4}.
\end{align}
While this is, at first glance, a promising explanation for the lack of thinner sheets producing ESEs, the above scalings were derived in the context of the MHD dynamo, which is not quantitatively applicable to turbulence in the ISM, where there is a mean magnetic field whose strength is comparable to that of the fluctuating magnetic field. However, the formation and disruption of current sheets may be qualitatively similar in the limit of large-amplitude turbulence, $\delta B / B_0 \gtrsim 1 $, which is more representative of typical ISM conditions, and in which the velocity field is also sufficiently strong to cause significant tangling and folding of field lines. By contrast, we note that in turbulence with a strong guide field ($\delta B/B_0 \lesssim 1$), the sheets created by dynamic alignment are expected to be disrupted on scales orders of magnitude smaller than the AU scales relevant for CR scattering and ESEs 
(\citealt{BL:2017}; \citealt{Mallet:2017}; \citealt{Schekochihin:2022}).

Tearing and turbulent decorrelation also determine the lifetime of a sheet. In order to produce long-duration ESEs, the sheets need to stay coherent for at least the duration of the ESEs, which typically ranges from months to years. The tearing timescale of $\sim$AU sheets is many orders of magnitude larger, with the growth rate of the fastest growing tearing mode scaling as $\gamma \sim ({\rm v_A} / s) \ S^{-3/7} {\rm Pm}^{-2/7}$, where the Alfv\'en-crossing time of the field reversal, $s/{\rm v_A}$, is of order a year, ${\rm Pm} \sim 10^{11}$ and the Lundquist number $S = s {\rm v_A}  / \eta \sim 10^{12}$ in the warm ISM. The turbulent lifetime of the sheet is in turn related to the rate at which it is being stretched along its long axis, $\sim \delta u(\ell)/\ell$ and the rate at which it is thinned along its short axis, $\sim \delta u(s) / s$. For a nearly incompressible flow, the two timescales are comparable and over 2 orders of magnitude longer than the duration of ESEs, assuming that $\delta u(s) \sim {\rm v_A} (s/L)^{1/3}$ as in Kolmogorov-like turbulence. Thus, the dynamical evolution of the sheets due to tearing and turbulent motions is slow enough to not decorrelate the structures on ESE timescales.

\section{Conclusions }\label{sec:conclusions}
In this work, we hypothesize that the same intermittent ISM structures may mediate cosmic ray (CR) propagation and give rise to extreme scattering events (ESEs) of radio waves. ESEs occur when the light from radio sources smaller in angular extent than milli-arcsec, e.g., pulsars and distant quasars, is strongly deflected by an underdense or overdense interstellar plasma lens, typically constrained to have a transverse size of order an AU. This scale is comparable to the gyroradius of a GeV CR in the Galaxy, which motivates our exploring of the connection between CR scattering and ESEs.  

The large ray deflection angles of ESEs require large gradients in the electron column density along the line of sight. Such large column density gradients can be explained by sheets of overdense/underdense plasma aligned with the line of sight, a model commonly adopted in the ESE and pulsar scintillation literature. In this paper, we have argued that density sheets that are kept in pressure balance by an enveloping reversing magnetic field (Section~\ref{sec:geometry} and Figure~\ref{fig:scattering_schematic}), are efficient scatterers of both radio waves (\citealt{2006GoldreichSridhar}) and CRs (\citealt{Lemoine:2023}; \citealt{Kempski:2023}). Such sheets are a generic product of theoretical and numerical models of the MHD turbulent dynamo (\citealt{Schekochihin:2004}), which suggests that they may be a product of large-amplitude interstellar turbulence. As illustrated in Figure~\ref{fig:scattering_schematic}, CR scattering and ESEs may be closely related; we indeed demonstrate that observationally derived CR scattering rates can be used to ``predict'' the ESE occurrence fraction and vice versa. 
We show that nearly straight magnetic/density sheets with axial ratios of order $10^4$, volume-filling factors of order $10^{-2}$ and electron density variations of order $0.03 \ {\rm cm}^{-3}$ (i.e., comparable to typical densities in the warm ISM) can simultaneously account for observed CR scattering rates and ESE observations (summarized in Section~\ref{sec:obs_constraints}). 

The two most demanding assumptions of the model are as follows. 1) We assume that current sheets remain straight for $10^4-10^5$ AU without bending by more than $10$ AU. Similar assumptions have been implicitly made by all the models that invoked sheets to explain ESEs and pulsar scintillations. Allowing for sheet curvature only further increases the demands on axial ratio, and makes CR scattering rates and ESE occurrence fractions less consistent. We briefly discuss physical processes that may set the geometry and morphology of the sheets in Section~\ref{sec:discussion}, but we encourage more rigorous numerical calculations of sheet morphologies found in simulations of plasma turbulence. 2) We assume that the statistics of sheets inferred from ESEs, which likely occur within the inner 1 kpc of Galactic latitudes, are representative of CR scattering across a much broader region (in particular, at larger heights above the Galactic disk). The validity of this assumption warrants further investigation.

The remarkable consistency that we find between the CR and ESE observables without the need of free parameters strongly motivates further investigation of the connection between the two phenomena. 
While direct observational constraints on GeV CR propagation are challenging to obtain, significant advancements are expected in the study of ESEs and pulsar scintillations over the next five years with the advent of wide-field radio surveys, such as CHORD and DSA-2000. These surveys may offer a novel tool for directly probing the interstellar structures responsible for CR scattering.
We propose investigating ESEs with pulsars, FRBs, and gravitationally lensed quasars to increase the statistical sample of sheets and better constrain their axial ratio. Additionally, we predict the existence of long-duration ESEs with low DM curves, and discuss the influence of these sheets on DM structure functions. With the large number of radio sources expected to be monitored in the upcoming surveys, we anticipate a significant improvement in our understanding of small-scale structures in the ISM, providing valuable constraints for simulations of interstellar turbulence and CR propagation models.

\begin{acknowledgments}
We thank Iryna Butsky, Phil Hopkins, Yuri Levin, Stephen Majeski, Ue-Li Pen, and Marten van Kerkwijk for many helpful comments. 
We sincerely appreciate the stimulating discussions at the Fields, Flows, and Filaments in the Magnetic ISM workshop, organized by Roger Blandford, Susan Clark, Roger Romani, and Dan Stinebring. 
We thank Lucy Reading-Ikkanda for help making Figure~\ref{fig:scattering_schematic}. 
We are grateful to Alisa Galishnikova for providing high-resolution snapshots of MHD dynamo fields (\citealt{Galishnikova:2022}), which were simulated as part of the Frontera computing project at the Texas Advanced Computing Center. The MHD turbulence simulation shown in Figure~\ref{fig:dBB4_slices} from \cite{Kempski:2023} was performed at facilities supported by the Scientific Computing Core at the Flatiron Institute, a division of the Simons Foundation. This research used resources of the Oak Ridge Leadership Computing Facility, which is a DOE Office of Science User Facility supported under Contract DE-AC05-00OR22725. 
EQ was supported in part by NSF AST Award No.~2107872 and by a Simons Investigator grant.
The research of ESP was supported in part by the
Gordon and Betty Moore Foundation through Grant
GBMF5076.
MWK was supported in part by NSF CAREER Award No.~1944972. 
AP acknowledges support by an Alfred P.~Sloan Research Fellowship and a Packard Foundation Fellowship in Science and Engineering. 
This research was supported in part by grant NSF PHY-2309135 to the Kavli Institute for Theoretical Physics (KITP), and performed in part at Aspen Center for Physics, which is supported by National Science Foundation grant PHY-2210452.
\end{acknowledgments}

\section*{Data Availability}
The calculations from this article will be shared on reasonable request to the corresponding author.



\appendix
\section{Lensing behavior seen in the Source Plane} \label{sec:gaussian_lens}
In this section, we provide a more complete lensing calculation that confirms~\eqref{eq:wsrcnumber} and shows the characteristic behavior of the lens that was used to constrain all the sheet parameters in the main text. 

The projection of an aligned sheet on the sky plane introduces a sudden change in the electron column density $N_e$ in the direction normal to the sheet (here, the $x$ direction). We use a Gaussian function to approximate the projected column density variation: 
\begin{equation}
    \Delta N_e = \Delta n_e\, l\, e^{-\frac{x^2}{2\sigma_s^2}}= \Delta N_{e,0}\, e^{-\frac{\theta^2}{2\sigma_\theta^2}},
\end{equation}
where $\sigma_s\approx s/2$. 
Following the notation used in gravitational lensing, we define the angle in the lens plane $\theta = x/D_\mr{lens}$ (Figure~\ref{fig:lensFreq}, and hence $\sigma_\theta = \sigma_s/D_\mr{lens}$. 

The variation of the electron column density introduce a chromatic dispersive potential,
\begin{align}
\Psi (\vec\theta) = \lambda^2 \frac{r_e}{2\pi} D_\mr{eff} \Delta N_e (\theta) 
=\sigma_\theta^2 \kappa e^{-\frac{\theta^2}{2\sigma_\theta^2}},
\end{align}
where $D_\mr{eff}=D_{ls}D_\mr{lens}/D_\mr{src}$; and the convergence $\kappa$ is given by
\begin{align}
\kappa &=  \lambda^2 \frac{r_e}{2\pi} D_\mr{eff} \Delta N_e^0 \sigma_s^{-2}
=20 \left(\frac{\nu}{\mr{GHz}}\right)^{-2}\left(\frac{D_\mr{eff}}{\mr{kpc}}\right)\left(\frac{\Delta N_{e,0}}{10^{16}\mr{cm}^{-2}}\right)\left(\frac{s}{\mr{AU}}\right)^{-2}.
\label{eq:kappa0}
\end{align}
For an underdense lens with $\Delta N_{e,0}<0$, $\kappa<0$; and for an overdense lens with $\Delta N_{e,0}>0$, $\kappa>0$. 

The gradient of the potential $\Psi (\theta)
$ determines the difference between the angular position of the source $\beta$ and the angular position of the lensed image  $\theta$: 
\begin{align}
\vec\beta  =  \vec\theta-\frac{{\rm D}_{\rm ls}}{{\rm D}_{\rm src}}
\vec{\hat\alpha}({\rm D}_{\rm d}\vec\theta)  =  \vec\theta-\vec\alpha(\vec\theta), \,\,
\vec\alpha(\vec\theta) = \frac{d}{d\vec \theta}{\Psi(\vec\theta)}.
\label{eq:deflect}
\end{align} 
The second gradient of the potential $\Psi (\theta)
$ determines the magnification of the image at $\theta$: $\mu=\left(1-\frac{d^2 \Psi}{d\theta^2}\right)^{-1}$. 
Caustics are observed when $d^2 \Psi/d\theta^2|_\mr{caustic}=1$. 

As shown in Figure~\ref{fig:lensFreq}, for an overdense lens, the source appears demagnified when in the region between the two inner caustics $\beta_\mr{demag}$. The position of the inner caustic depends weakly on $\kappa$, and hence frequency, such that 
\begin{equation}
\beta_\mr{demag}\approx \pm 3 \sigma_\theta.
\label{eq:betademag}
\end{equation}
Therefore, the duration of the demagnification can be used to estimate the width of the lens, i.e. the thickness of the sheet. 

The area between the two outer caustic can be used to approximate the area affected by the aligned sheet.  We can solve for the position of the outer caustics $\theta_\mr{outer,\pm}=\pm (1+\sqrt{e}/2\kappa)\sigma_\theta$. And from Eq.~\ref{eq:deflect}, we can obtain $\beta_\mr{focal,\pm}=\theta_\mr{focal}(1-\sigma_\theta^2/\theta_\mr{focal}^2)$. When $\kappa<2.24$, no caustics are formed. In the limit $\kappa\gg 1$: 
\begin{equation}
    \beta_\mr{focal,\pm}=\pm \frac{\kappa}{\sqrt{e}}\sigma_\theta.
    \label{eq:betaout}
\end{equation}
The lensing is negligible in the $y$ direction, $\theta_y\approx \beta_y$, hence: 
\begin{equation}
    \frac{\Omega_\mr{focal}}{\Omega_\mr{lens}}\approx \frac{\beta_\mr{focal,+}-\beta_\mr{focal,-}}{2\sigma_\mr{\theta}} \approx \kappa.
    \label{eq:lens2src}
\end{equation}
Similar results can be obtained for underdense lenses. Two caustics are formed when $\kappa<-1$ and the area between them can be used to approximate the affected area in the source plane. The relation $\beta_\mr{focal,\pm}=\pm \kappa \sigma_\theta/\sqrt{e}$ remains true for $|\kappa| \gg 1$. Therefore, equation~\ref{eq:lens2src} also holds in the case of an underdense lens. Thus, we can derive the chromatic occurrence fraction of a radio source being plasma lensed by an aligned sheet from Eqs. ~\ref{eq:wl},~\ref{eq:wsrc2lens},~\ref{eq:kappa0} and~\ref{eq:lens2src}: 
\begin{align}
 W_\mr{src} &= \frac{L f_V s}{\ell^2} \kappa =4\lambda^2 \frac{r_e}{2\pi} D_\mr{eff} L \Delta n_{e,0}
 \frac{1}{s} \frac{f_V}{l}.
 \label{eq:wsrc}
 \end{align}

\section{Alternative models of scattering in sheets} \label{sec:alternative_sheet} 
The scattering model illustrated by the schematic in Figure~\ref{fig:scattering_schematic} assumes magnetic-field lines that envelope the density sheet in an aligned way. While this is motivated by the theory of how magnetic fields are amplified by turbulent flows (\citealt{Schekochihin:2004}; \citealt{Rincon:2019}), it is nevertheless worth considering alternative models of field-line morphology around density sheets. For example, we can consider the case in which CR scattering occurs in the entire volume of the sheet and magnetic-field lines enter the sheet from random directions. Thus, instead of CR scattering by bends at the ends of the sheets as in Figure~\ref{fig:scattering_schematic}, we now consider CR scattering regions to be exactly co-spatial with the density sheet. This could for example be the case if sheets of high magnetic field-line curvature are associated with sheets of large density gradients, which could, e.g., occur at an oblique shock interface. In this case, the CR mean free path is smaller than predicted by~\eqref{eq:cr_mfp_fold} and is given by (\citealt{Butsky:2024}),
\begin{equation}
    \lambda_{\rm CR} \sim \frac{s}{f_V},
\end{equation}
where $s$ is the thickness of high-curvature/density sheet and $f_V$ is the volume-filling factor of the sheets. The ESE mean free path is unchanged, and so for straight sheets (cf. eq.~\ref{eq:lambda_pl_cr}),
\begin{equation}
    \lambda_{\rm PL, straight} \sim \lambda_{\rm CR} \Big(\frac{\ell}{s}\Big)^2.
\end{equation}
Thus, for straight sheets with $\ell/s \sim 10^4$ to reproduce the large bending angles in  ESEs, the distance a radio wave has to travel to encounter an aligned sheet is $\sim 10^8$ times larger than the CR mean free path, a value much larger than constrained by observations. However, if the sheet has a significant global radius of curvature $r_c$, its alignment probability is $\sim r_c/\ell$ instead of $s / \ell$ and so for $r_c \sim \ell$,
\begin{equation}
    \lambda_{\rm PL, curved} \sim \lambda_{\rm CR} \frac{\ell}{s}.
\end{equation}
This is similar to the result for straight sheets in the model considered in Figure~\ref{fig:scattering_schematic} and Section~\ref{sec:chance_align}, where we found that CR scattering and ESE rates were roughly compatible for $\ell / s \sim 10^4$. Thus, the alternative model considered here may result in roughly compatible CR scattering and ESE rates for sheets with significant global radii of curvature, a significant shortcoming of the model from Figure~\ref{fig:scattering_schematic}. However, while for straight sheets $\ell / s \sim 10^4$ is sufficient to explain the large ESE bending angles, for $r_c \sim \ell$ the bending angle is a 100 times smaller at fixed $\Delta n_e$, as the maximum length a radio wave can travel along the sheet is $(r_c s)^{1/2}$. Therefore, it underpredicts ESE bending angles unless the density perturbations are much larger than the typical density variance in the warm ISM.

\section{Quasi-spherical structures} \label{sec:spherical_geometry}
Quasi-spherical structures are unlikely sites of ESEs owing to the fact that they require extreme number densities of ionized gas $\sim 1000 \ {\rm cm^{-3}}$ to produce the observed large bending angles. Such high number densities would be highly overpressurized relative to the surrounding ISM and so these structures would quickly disperse. Here we nevertheless set aside this issue and briefly consider whether quasi-spherical structures could simultaneously account for the observed rates of ESEs and CR scattering.

Both for radio waves undergoing ESEs and $\sim$GeV CRs, the mean free path due to intermittent scattering off spherical AU-scale structures is of order $(n_s A)^{-1}$, where $n_s$ is the number density of structures and $A$ is their cross section. However, the observed occurrence fraction of ESEs ($\sim 10^{-3} - 10^{-2}$) corresponds to an ESE mean free path of at least 100 kpc assuming a 1 kpc sightline through the Milky Way's warm ISM. This is drastically larger (by more than 4 orders of magnitude) than the observed CR mean free path, which is of order $1-10$ pc. In the presence of magnetic focusing, i.e. if the magnetic field in the lens $B_{\rm lens}$ is larger than typical ISM fields $B_{\rm ISM}$, the effective cross section for CRs is larger by a factor $B_{\rm lens}/B_{\rm ISM}$. Thus, the CR mean free path would be reduced relative to the ESE mean free path. However, consistency with observations then requires that the lensing structures host strong magnetic fields exceeding $\sim 10 \  {\rm mG}$, which would in turn further add to the overpressure problem. Thus, it is highly unlikely that quasi-spherical structures can simultaneously account for the observed ESE and CR scattering rates. 


\bibliography{ese,cr}{}
\bibliographystyle{aasjournal}



\end{document}